\documentclass[fleqn,10pt]{wlscirep}

\usepackage{setspace}
\usepackage{amsmath}
\usepackage{amssymb} 
\usepackage{mathrsfs} 
\usepackage{fixltx2e}
\usepackage{booktabs} 
\usepackage{todonotes}
\usepackage{multirow}
\usepackage{graphicx,psfrag,tikz}
\usepackage{color}
\usepackage{pdflscape}		
\usepackage{caption}
\usepackage{subcaption}
\usepackage[normalem]{ulem}           
\usepackage{fp} 
\usepackage{booktabs} 
\usepackage[numbers]{natbib}

\newcommand{\FFRd}[1]{FFR\textsubscript{#1D}}
\newcommand{\FFR}{FFR }
\newcommand{\Reynolds}{\text{Re}}

\usetikzlibrary{arrows}

\title{Comparison of 1D and 3D Models for the Estimation of Fractional Flow Reserve}

\author[1,3,*]{P.J.~Blanco}
\author[1,3]{C.A.~Bulant}
\author[1,3]{L.O.~M\"uller}
\author[1,3]{G.D.~Maso Talou}
\author[2,3]{C.~Guedes Bezerra}
\author[2,3]{P.L.~Lemos}
\author[1,3]{R.A.~Feij\'oo}

\affil[*]{pjblanco@lncc.br}
\affil[1]{National Laboratory for Scientific Computing, LNCC/MCTIC, Av. Get\'ulio Vargas, 333, Petr\'opolis-RJ, 25651-075, Brazil}
\affil[2]{Department of Interventional Cardiology, Heart Institute (InCor) and the University of S\~ao Paulo Medical School, S\~ao Paulo, SP, 05403-904, Brazil}
\affil[3]{INCT-MACC Instituto Nacional de Ci\^encia e Tecnologia em Medicina Assistida por Computa\c c\~ao Cient\'ifica, Petr\'{o}polis, Brazil}

\begin{abstract}
In this work we propose to validate the predictive capabilities of one-dimensional (1D) blood flow models with full three-dimensional (3D) models in the context of patient-specific coronary hemodynamics in hyperemic conditions. Such conditions mimic the state of coronary circulation during the acquisition of the Fractional Flow Reserve (FFR) index. Demonstrating that 1D models accurately reproduce FFR estimates obtained with 3D models has implications in the approach to computationally estimate FFR. To this end, a sample of 20 patients was employed from which 29 3D geometries of arterial trees were constructed, 9 obtained from coronary computed tomography angiography (CCTA) and 20 from intra-vascular ultrasound (IVUS). For each 3D arterial model, a 1D counterpart was generated. The same outflow and inlet pressure boundary conditions were applied to both (3D and 1D) models.
In the 1D setting, pressure losses at stenoses and bifurcations were accounted for through specific lumped models. Comparisons between 1D models (\FFRd{1}) and 3D models (\FFRd{3}) were performed in terms of predicted $\text{FFR}$ value.
Compared to \FFRd{3}, \FFRd{1} resulted with a difference of 0.00$\pm$0.03 and overall predictive capability AUC, Acc, Spe, Sen, PPV and NPV of 0.97, 0.98, 0.90, 0.99, 0.82, and 0.99, with an FFR threshold of 0.8. We conclude that inexpensive \FFRd{1} simulations can be reliably used as a surrogate of demanding \FFRd{3} computations.
\end{abstract}


\begin{document}

\flushbottom
\maketitle

\thispagestyle{empty}

\section*{Introduction}

Fractional Flow Reserve (FFR) is a hemodynamic index aimed at the quantification of 
the functional severity of a coronary artery stenosis. This index, which is calculated from pressure measurements 
and under hyperemic conditions, has been proposed and used to detect myocardial ischemia 
\cite{pijls_measurement_1996,de_bruyne_pressure-derived_2000}, and has largely demonstrated excellent results as a 
diagnostic tool to defer patients with intermediate lesions to surgical procedures \cite{pijls_percutaneous_2007,Tonino_FFR_FAME-2009,vanNunen_FFR_FAME5-2015}.

Making use of valuable information regarding anatomy and vascular geometry contained in medical images, 
the scientific community specialized in computational models initiated a race pursuing the paradigm of noninvasive estimation of FFR 
through the use of computer simulations of coronary blood flow. A myriad of different approaches 
using image modalities such as coronary computed tomography angiography (CCTA) \cite{taylor_computational_2013}, angiography (AX) \cite{morris_virtual_2013}, and even optical coherence tomography (OCT) \cite{Ha_FFRoct-2016}, emerged. These approaches employ 3D models to estimate pressure losses in coronary vessels and thus to devise a strategy to predict patient-specific FFR.
Such technology has been useful to improve diagnostic accuracy with respect to different traditional protocols taking the invasive measurement of FFR as gold standard \cite{koo_diagnosis_2011,min_diagnostic_2012,yoon_noninvasive_2012}.
It is important to remark that the use of 3D models carry several challenges, which range from detailed 3D lumen segmentation procedures and mesh generation to time-consuming numerical simulations in high performance computing facilities.

In turn, simplified mathematical models, either based on the 1D Navier-Stokes equations in compliant vessels \cite{huo_hybrid_2007} or 
based on compartmental (0D) representations \cite{maasrani_analog_2008} have been employed to study different aspects of coronary physiology.
Since these models neglect fundamental aspects of the 3D physics regarding flow across geometric singularities, specific models to account 
for focal pressure losses are usually employed \cite{young_1973,huo_validated_2012,mynard_2015}. However, these lumped models require the definition of 
specific parameters which may result in a cumbersome task.
Validation of 1D models by using 3D simulations as gold standard for idealized phantoms and patient-specific arterial districts such as the cerebral arteries, the aorta and major vessels were reported elsewhere \cite{grinberg_modeling_2011,jonasova_comparative_2014,xiao_systematic_2014,alastruey_impact_2016}.
However, no validation was performed combining hyperemia, stenotic lesions and patient specific coronary territories, which are key ingredients in the computational estimation of FFR.
Yet the use of 1D models to estimate FFR using CCTA or AX images has been recently proposed \cite{Renker_FFRSiemens-2014,Coenen_FFRSiemens-2014,Trobs_FFRSiemensAngio-2016,Ko_FFRToshibaCTFFR-2016}.
It is worth noting that, in the context of coronary circulation, the validation of 1D models against 3D models has only been reported in \cite{Boileau_FFR1D-2017}, where a virtual patient population derived from a single CT scan was used. However, a comprehensive validation that accounts for large variability observed in terms of lumen geometry and hyperemic flow conditions is lacking.

The goal of this work is to demonstrate that 1D models are capable of predicting FFR (denoted \FFRd{1}) with a high degree of accuracy when compared 
with FFR predicted by 3D models (denoted \FFRd{3}), which is used as ground truth. To achieve this goal we make use of a sample of patients for which CCTA and IVUS images were available.
After the generation of 3D models, the corresponding 1D centerlines representing the arterial topology were extracted. Then, 3D and 1D numerical simulations were performed under the assumption of hyperemic flow conditions and using appropriate boundary conditions.
Several modeling scenarios are proposed from which two stand out: a scenario for practical use, and a best case scenario. The former scenario contains generic model parameters which need no tuning whatsoever. The later scenario represents the ultimate 1D modeling approach, in which several model parameters have been estimated to match data extracted from 3D models.


\section*{Material and Methods}\label{sec:Methods}

\subsection*{Vascular Models}\label{sec:Methods.VascularModels}

The patient sample consisted of 20 patients who were referred to CCTA and/or IVUS protocols for diagnostic or therapeutic percutaneous coronary procedure at S\'irio-Liban\^es Hospital, S\~ao Paulo, Brazil. A total of 29 reconstructed arterial trees were available, 9 obtained from coronary computed tomography angiography (CCTA) and 20 constructed from intra-vascular ultrasound (IVUS).

CCTA images were acquired at end-diastole. Segmentation of arterial geometries from CCTA was performed using implicit deformable 
models \cite{Antiga_CollidingFronts-2008} with tools available in the vmtk library \cite{vmtkProject}. 
IVUS datasets contain IVUS runs and angiographic images (AX), which were synchronized with the ECG signal during the acquisition. 
End-diastolic frames are gated \cite{Maso_Talou2015} and registered \cite{Maso_Talou2017} to construct the 3D geometry using deformable models \cite{MasoTalou2013}.
Branches in IVUS models were manually segmented and added to the vascular model. 
See \cite{bulant_head_2016} for details on the image and mesh processing for each imaging modality.

\begin{figure}[ht]
\centering
\begin{tikzpicture}[scale=0.80,transform shape]
\node at (-3*\xjump,0) [] {\includegraphics[height=\size cm]{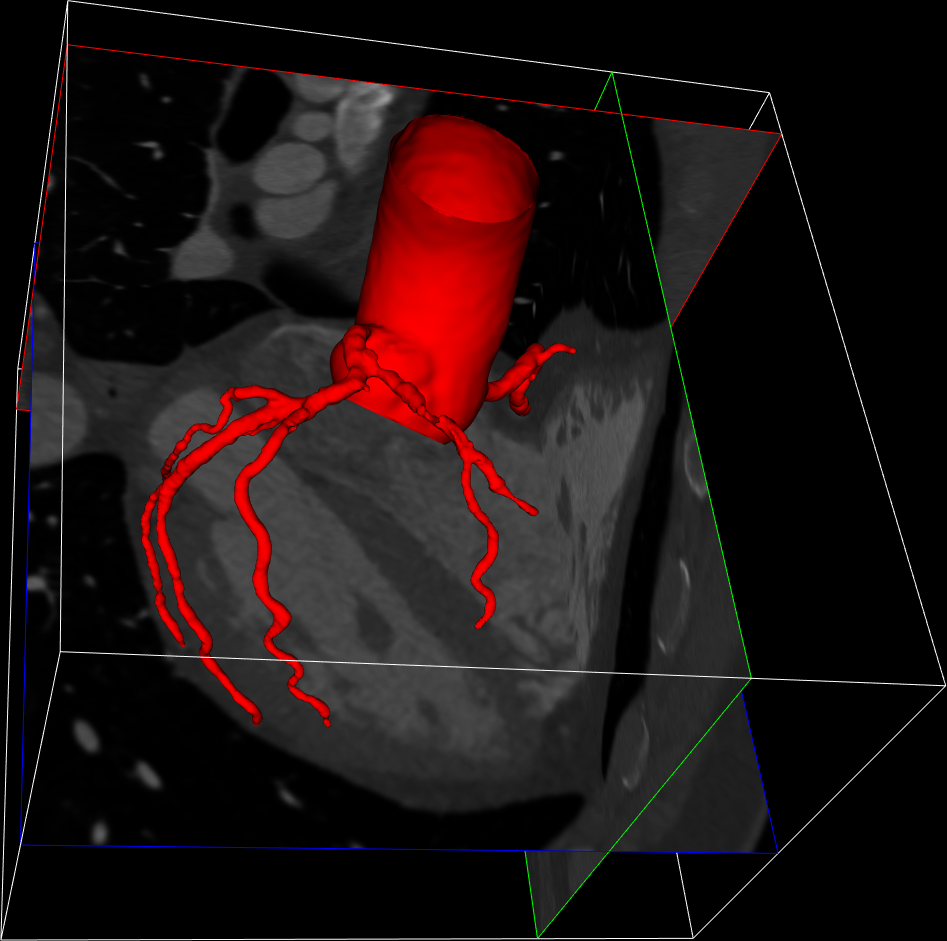}};
\node[font=\scriptsize] at (-3*\xjump,-\yjump) [] {Image processing};
\node at (-3*\xjump,-3*\yjump) [] {\includegraphics[height=\size cm]{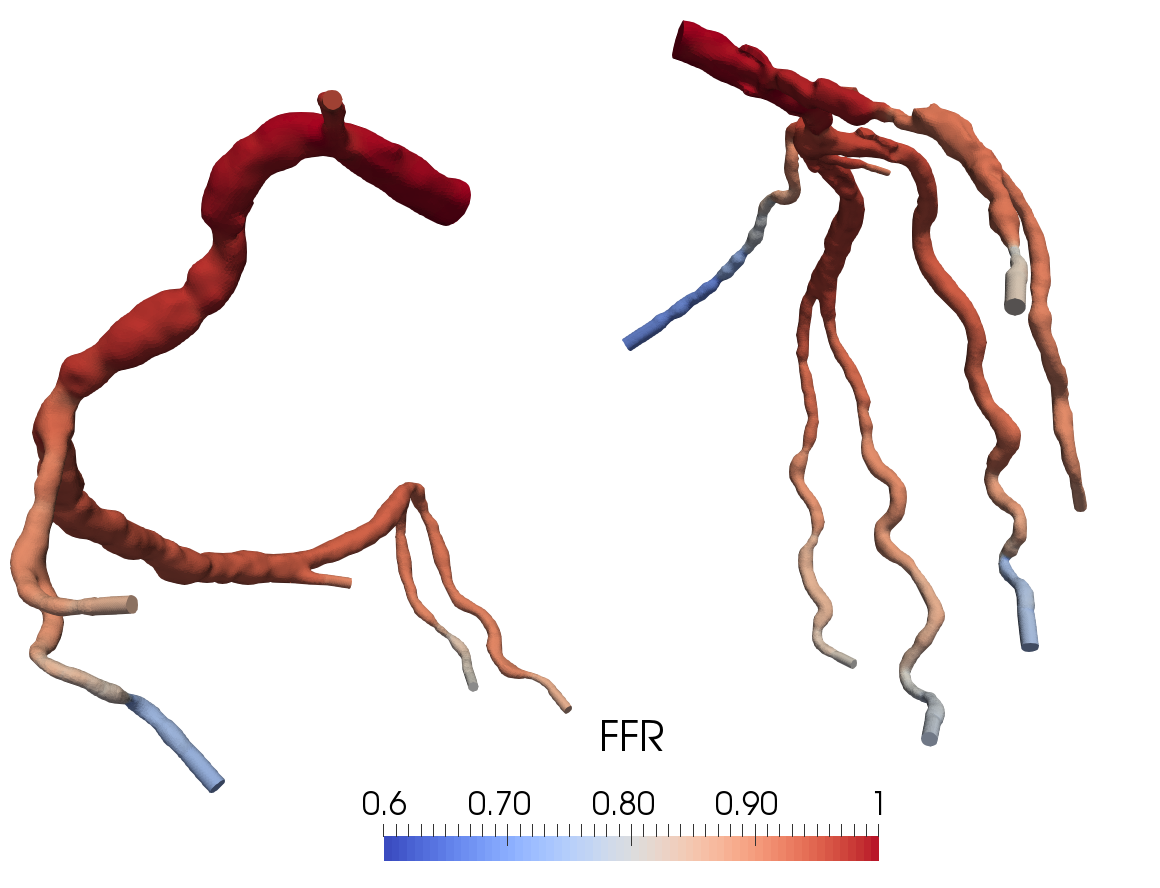}};
\node[font=\scriptsize] at (-3*\xjump,-4.25*\yjump) [] {3D simulation} ;
\draw[rounded corners,thick,red!50] (-4*\xjump,-4*\yjump) rectangle (-2*\xjump,-2*\yjump);

\node at (-\xjump,0) [] {\includegraphics[height=\size cm]{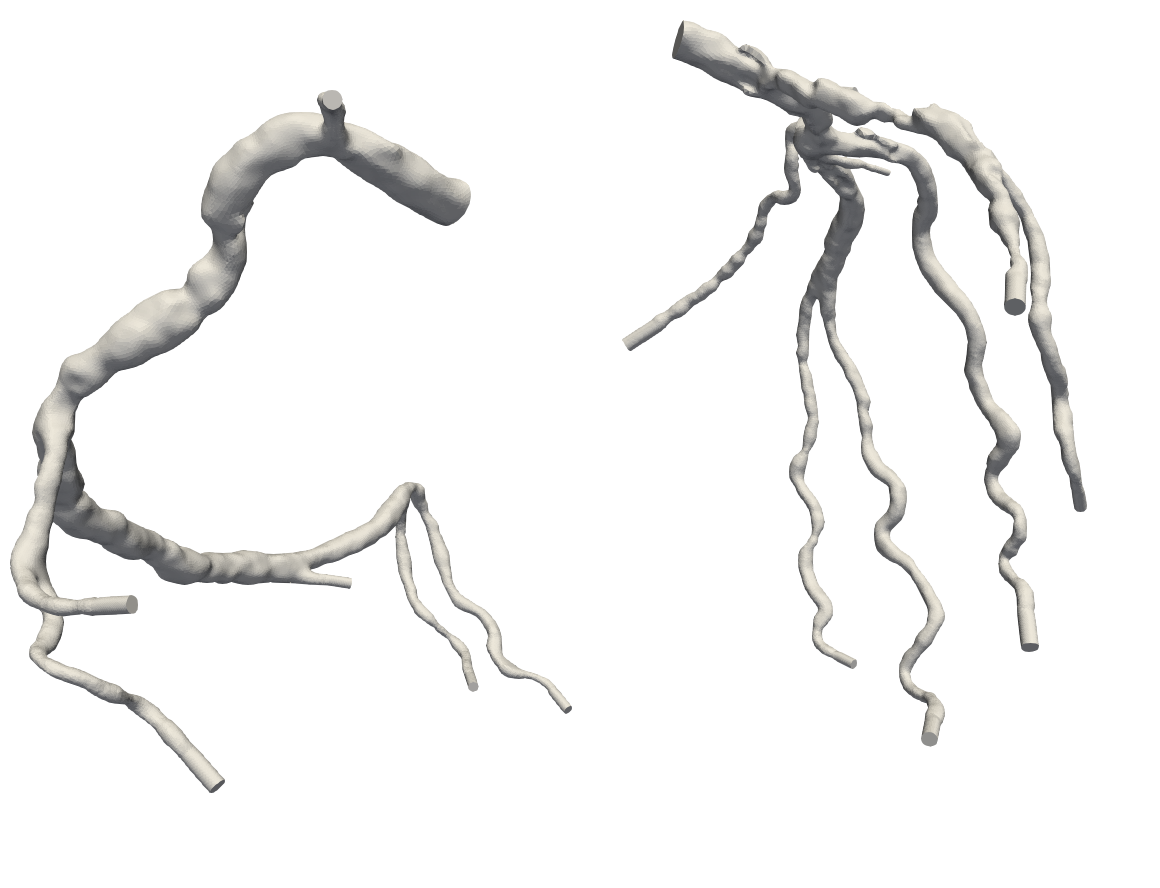}};
\node[font=\scriptsize] at (-\xjump,-\yjump) [] {Vascular (3D) model}; 
\node at (\xjump,0) [] {\includegraphics[height=\size cm]{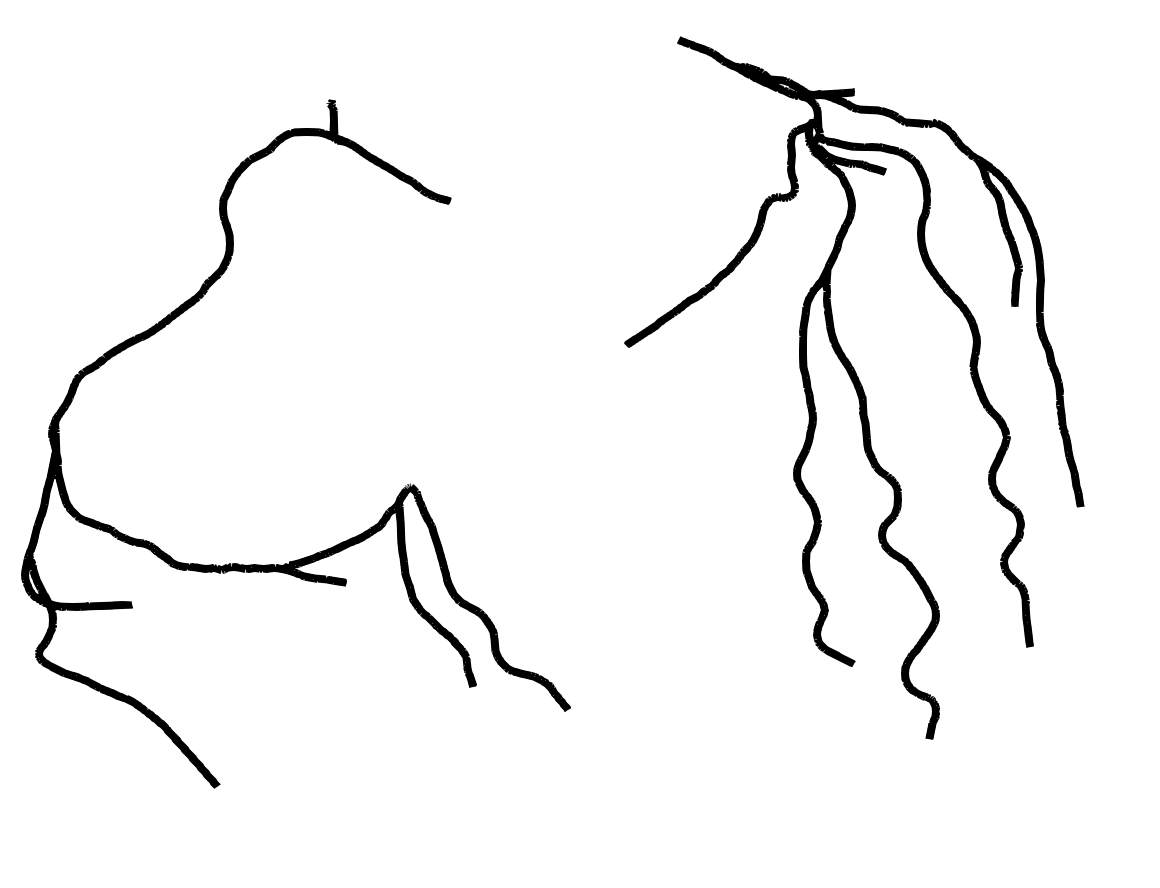}};
\node[font=\scriptsize] at (\xjump,-\yjump) [] {Centerline (1D) model};
\node at (3*\xjump,0) [] {\includegraphics[height=\size cm]{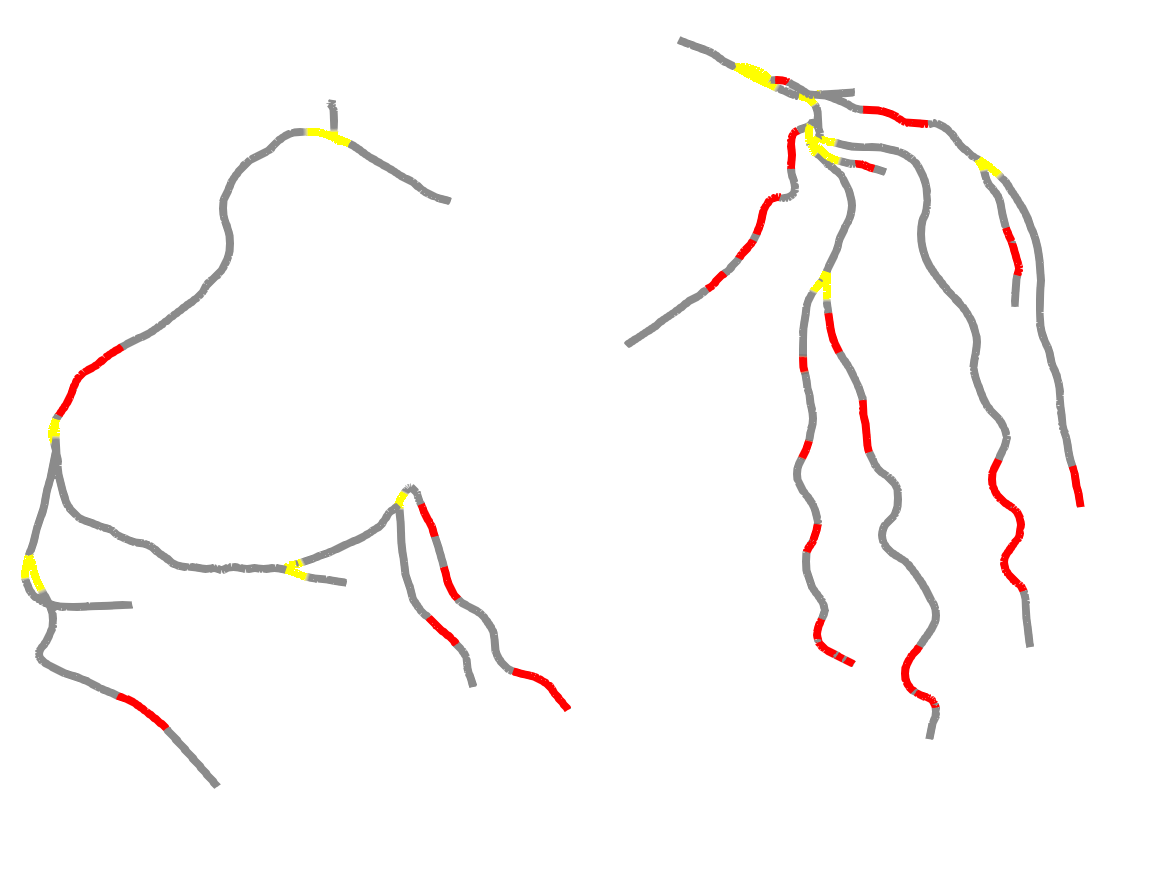}};
\node[font=\scriptsize] at (3*\xjump,-\yjump) [] {Centerline masks};

\draw[rounded corners,thick,red!50] (0,-1.2*\yjump) -- (-4*\xjump,-1.2*\yjump) -- (-4*\xjump,\yjump) -- (0,\yjump);
\draw[rounded corners,thick,blue!50] (0,\yjump) -- (4*\xjump,\yjump) -- (4*\xjump,-1.2*\yjump) -- (0,-1.2*\yjump);

\draw[o-stealth,thick,red!50] (-\xjump,-1.2*\yjump) -- (-\xjump,-1.5*\yjump) -- (-3*\xjump,-1.5*\yjump) -- (-3*\xjump,-2*\yjump);

\draw[o-stealth,thick,blue!50] (3*\xjump,-1.2*\yjump) -- (3*\xjump,-1.5*\yjump) -- (2.25*\xjump,-1.5*\yjump) -- (2.25*\xjump,-1.85*\yjump);

\draw[o-stealth,thick,orange!80] (-2*\xjump,-3*\yjump) -- (-1.7*\xjump,-3*\yjump) -- (-1.7*\xjump,-4.15*\yjump) -- (0.35*\xjump,-4.15*\yjump);
\node at (-0.7*\xjump,-3.85*\yjump) {{\scriptsize\begin{tabular}{c}
                               Data assimilation\\
                               $\Delta p \rightarrow f_K$\\
                               $p_i,i=1,\ldots,N_o \rightarrow f_\varpi$
                              \end{tabular}}};
\node at (-0.7*\xjump,-4.3*\yjump) {{\scriptsize verification purposes}};
\begin{scope}[yshift=-0.5*\yjump cm]
\node at (1.25*\xjump,-2*\yjump) [] {\includegraphics[height=\sizesm cm]{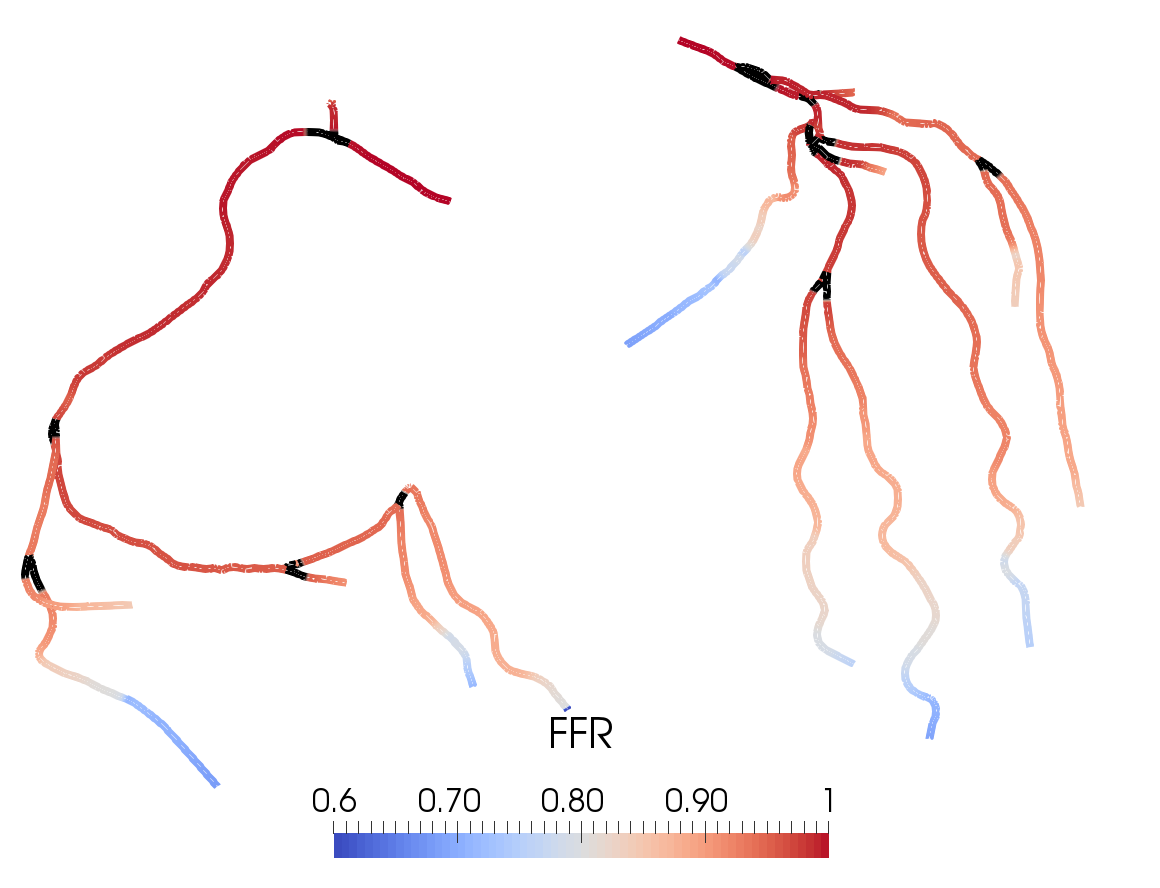}};
\node at (1.25*\xjump,-2.75*\yjump) [] {{\scriptsize R: $\varpi^o$}};
\node at (3.25*\xjump,-2*\yjump) [] {\includegraphics[height=\sizesm cm]{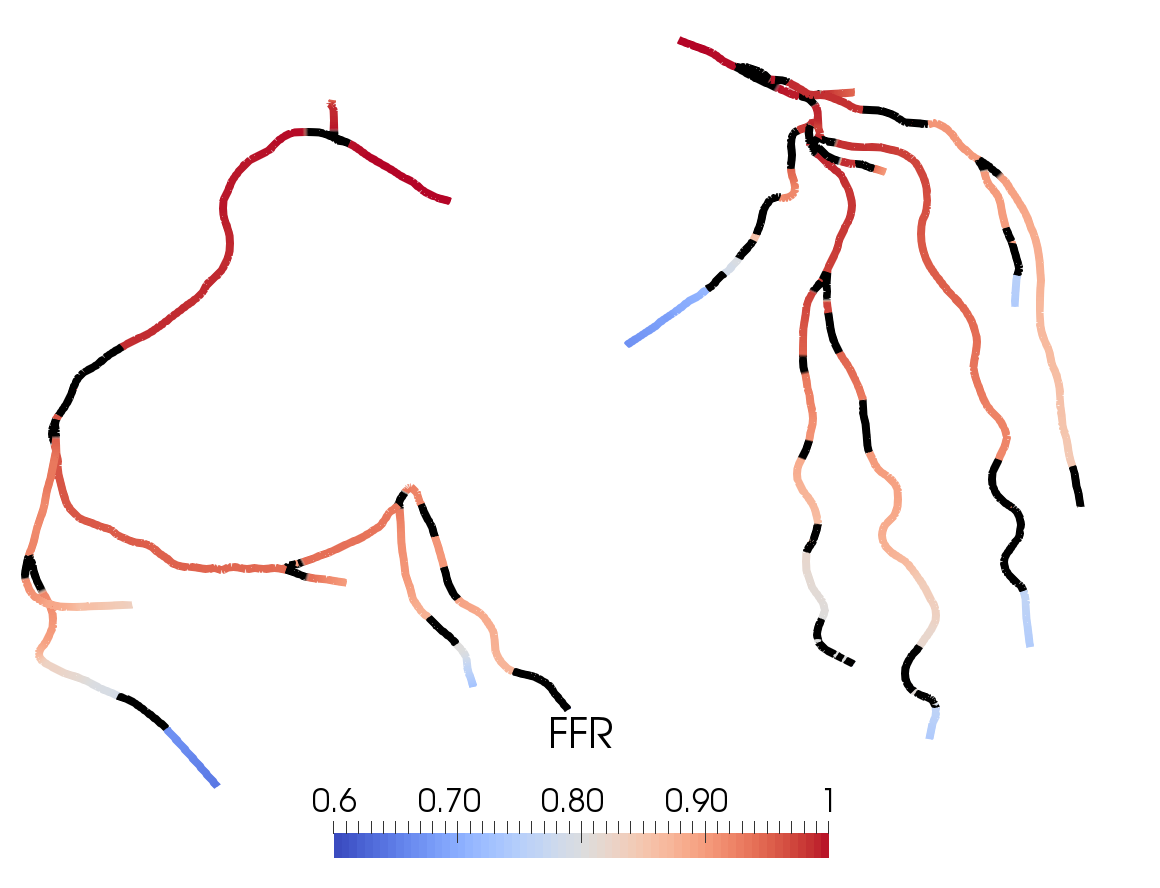}};
\node at (3.25*\xjump,-2.75*\yjump) [] {{\scriptsize P: $\varpi^o$, $K_v^o$}};

\node at (1.25*\xjump,-3.5*\yjump) [] {\includegraphics[height=\sizesm cm]{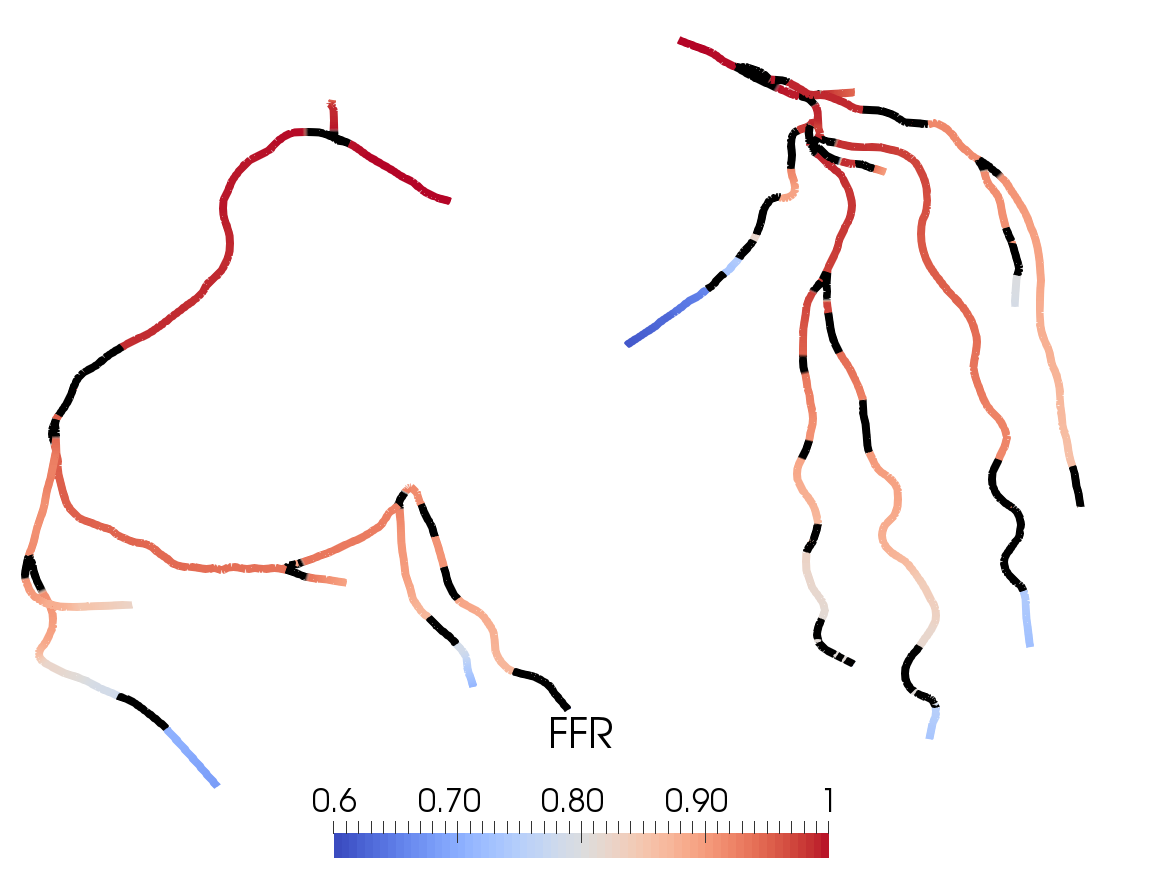}};
\node at (1.25*\xjump,-4.25*\yjump) [] {{\scriptsize I: $\varpi^o$, $K_v=f_KK_v^o$}};
\node at (3.25*\xjump,-3.5*\yjump) [] {\includegraphics[height=\sizesm cm]{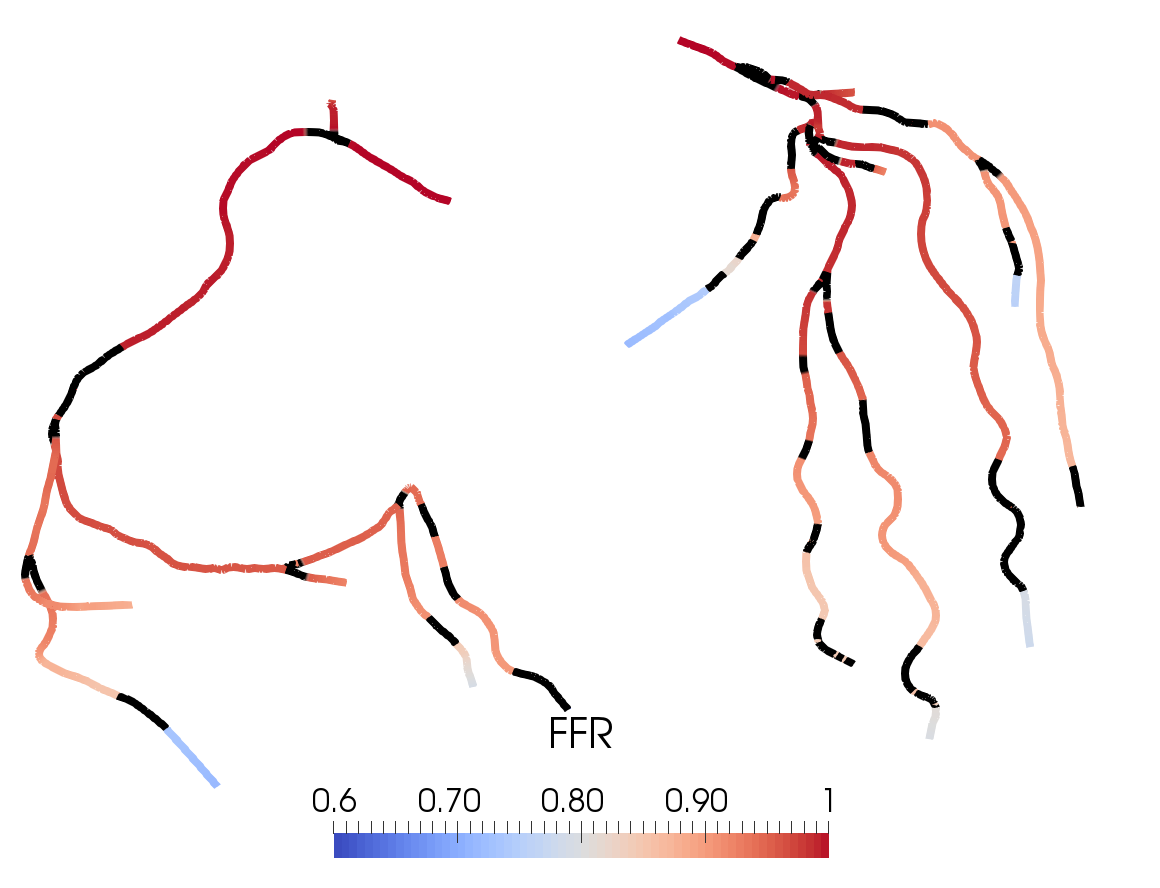}};
\node at (3.25*\xjump,-4.25*\yjump) [] {{\scriptsize B: $\varpi=f_\varpi\varpi^o$, $K_v=f_KK_v^o$}};
\draw[rounded corners,thick,blue!50] (0.25*\xjump,-4.5*\yjump) rectangle (4.25*\xjump,-1.35*\yjump);
\node[font=\scriptsize] at (2.25*\xjump,-4.75*\yjump) [] {1D simulations (R, P, I, B scenarios)};
\draw[rounded corners,dashed,orange!80,thick] (0.35*\xjump,-4.35*\yjump) rectangle (4.15*\xjump,-2.95*\yjump);
\end{scope}

\end{tikzpicture}
\caption{Workflow for the construction of vascular models. Image segmentation produces 3D vascular geometries which are processed to retrieve the 1D centerline geometry. Lumen area is given at each centerline point and bifurcation (yellow) and stenoses (red) masks are applied when necessary. Model scenarios are: R (\textit{raw}): 1D model with no stenoses and known dissipation parameter $\varpi^o$, P (\textit{practical}): idem R scenario,  but 1D model includes stenoses with known stenosis parameter $K_v^o$, I (\textit{intermediate}): idem P scenario, but parameters $K_v$ are estimated using stenosis drop pressures $\Delta p$ at the corresponding lesions from 3D simulations, B (\textit{best-case}): idem I scenario, but dissipation parameter $\varpi$ is estimated using outlet pressures $p_i,i=1,\ldots,N_o$ from 3D simulations.}\label{fig:1D3Dworkflow}
\end{figure}
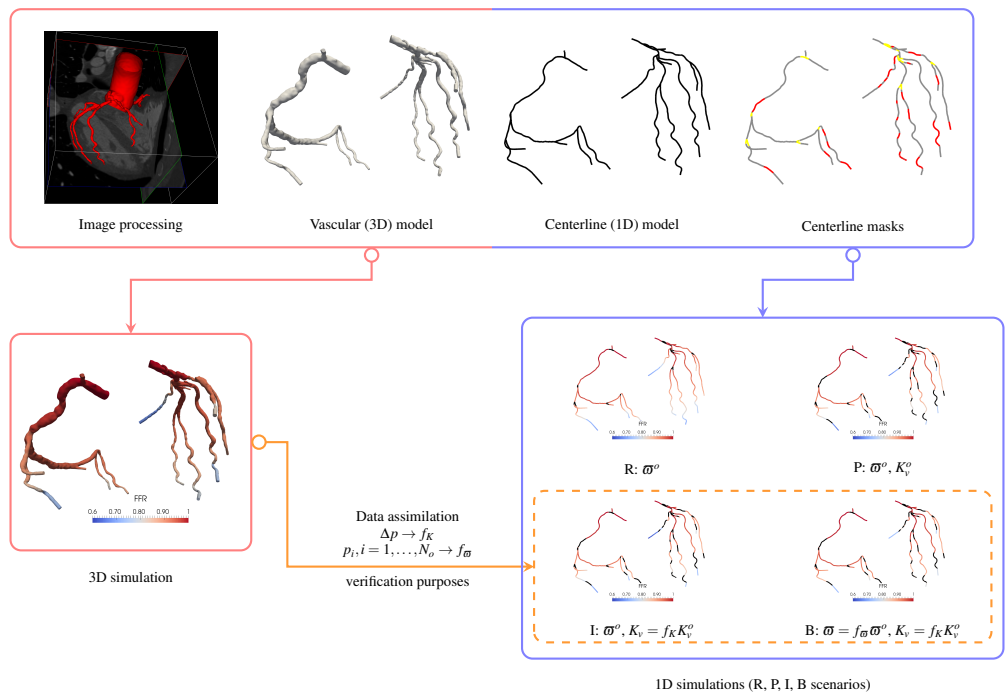

For 3D simulations, tetrahedral finite element meshes were built using the vmtk library \cite{vmtkProject}.
For 1D simulations, the centerline of each vascular model (resolution of $0.05\text{ cm}$ between points) was extracted \cite{Antiga_CenterLine-2003}. 
Each point of the centerline features a cross-sectional area in correspondence with the 3D model, from which the equivalent lumen radius is computed.
Figure~\ref{fig:1D3Dworkflow} presents the workflow for constructing the 3D and the 1D models.

The clinically interrogated vessel was known for each patient, as well as the position of a hypothetical invasive \FFR measurement.

\subsection{3D Models}\label{sec:Methods.3DModel}

Consider a rigid vascular domain $\Omega\in\mathbb R^3$, with boundary $\Gamma$ 
whose outward unit normal vector is $\mathbf n$. The boundary is decomposed into the inlet
boundary $\Gamma_i$, the lateral wall boundary $\Gamma_w$ and the $N_o$ outlets of the domain $\Gamma_o^k$, $k=1,\ldots,N_o$.
Blood flow is modeled as a Newtonian fluid, therefore, the Navier-Stokes equations for incompressible flows hold
\begin{equation}
\begin{cases}
\rho\frac{\partial\mathbf v}{\partial t}+\rho(\nabla\mathbf v)\mathbf v+\nabla p-\mu\triangle\mathbf v=\mathbf 0 & \text{in }\Omega,\\
\operatorname{div}\mathbf v=0&\text{in }\Omega,\\
\mathbf v=\mathbf 0&\text{on }\Gamma_w,\\
-p\mathbf n+2\mu(\nabla\mathbf v)^s\mathbf n=P_{\text{ao}}\mathbf n&\text{on }\Gamma_i,\\
-p\mathbf n+2\mu(\nabla\mathbf v)^s\mathbf n=P_{\text{out}}^k\mathbf n&\text{on }\Gamma_o^k,\\
\text{with }P_{\text{out}}^k=P_{\text{ref}}+R_{\text{out}}^k\int_{\Gamma_o^k}\mathbf v\cdot\mathbf n d\Gamma & k=1,\ldots,N_o.
\end{cases}
\end{equation}
where $\mathbf v$ and $p$ are the velocity and pressure fields, $(\cdot)^s$ denotes the symmetrization operation, and $\rho$ and $\mu$ are the 
fluid density and dynamic viscosity, respectively.
At the inlet, a reference aortic pressure $P_{\text{ao}}$ is prescribed as a Neumann boundary condition.
At the outlet, resistances $R_{\text{out}}^k$ are used to simulate the pressure losses in the remaining vasculature, up to 
a reference venous pressure $P_{\text{ref}}$, and are computed as explained in \cite{bulant_head_2016}, where the 3D finite element strategy used to find approximate solutions to this model is also described.

The flow rates through the different outlets delivered by the 3D simulation result
\begin{equation}
Q_{\text{out}}^k=\int_{\Gamma_o^k}\mathbf v\cdot\mathbf n d\Gamma\qquad k=1,\ldots,N_o,\label{eq:Q3D_out}
\end{equation}
which will be later used as boundary conditions for the 1D model.

\subsection{1D Models}\label{sec:Methods.1DModel}

\subsubsection{Mathematical Formulation}

Skeletonization of $\Omega$ yields a set of centerlines, representing arterial segments, connected through a set of junctions.
Centerline coordinate is denoted by $x$. The inlet boundary $\Gamma_i$ is now simply denoted by the inlet point $I$, and the outlet 
boundaries $\Gamma_o^k$ are denoted by $O_k$, $k=1,\ldots,N_o$.
Given a generic centerline of size $[0,L]$, the governing 1D equations are the following
\begin{equation}
\begin{cases}
\frac{\partial A}{\partial t}+\frac{\partial Q}{\partial x}=0&\text{in }[0,L],\\
\frac{\partial Q}{\partial t}+\frac{\partial}{\partial x}\big(\frac{Q^2}{A}\big)+\frac{A}{\rho}\frac{\partial P}{\partial x}+\frac{2 \varpi\pi\mu U}{\rho}=0&\text{in }[0,L],\\
P=P_0 + \beta\big(\sqrt{\frac{A}{A_0}}-1\big)&\text{in }[0,L],\\
\end{cases}\label{eq:1Dmodel}
\end{equation}
where $Q$ is the flow rate, $A$ is the lumen cross-sectional area, $P$ is the average pressure in the lumen cross section, $U=\frac{Q}{A}$ is the cross-sectional average velocity, $\varpi$ is a parameter that characterizes the velocity profile in the 1D model, $\beta$ is an effective stiffness which characterizes the compliance of the arterial wall, $P_0$ is a reference external pressure for which the lumen area is $A_0$. Since 1D models are to be compared with rigid wall 3D models, $\beta$ is set at a high value to mimick the 1D blood flow in a quasi-rigid domain (therefore $A\approx A_0$). In practice, a per-case estimation of $\beta$ was performed to ensure that maximum area deviations with respect to $A_0$ are smaller than $1\%$.

\subsubsection{Boundary Conditions}

Regarding boundary conditions, the inlet pressure in the 1D model is $P_{\text{ao}}$, the same as in the 3D model. For the outflow boundary conditions, the flow rates $Q_{\text{out}}^k$, $k=1,\ldots,N_o$, which are determined from \eqref{eq:Q3D_out}, are prescribed. In this manner, since the blood flow through the corresponding 3D and 1D outlets is exactly the same, we ensure the same flow distribution, and, therefore, it is possible to assess the predictive capabilities of the 1D model in determining the pressure drop along the coronary tree, and in particular at lesions, by direct comparison with the 3D model, which is regarded as the ground truth.

\subsubsection{Junction Models}

At junctions we consider mass conservation
\begin{equation}
\sum_{i=1}^{N_j}Q_i=0, \label{eq:masscons}
\end{equation}
where $N_j$ is the number of converging segment. For the remaining conservation equation at junctions we consider two models.
The first model is the \textit{standard junction model} (S model), and consists of the standard assumption of conservation of total pressure, that is
\begin{equation}
P_1+\frac{\rho}{2}U_1^2=P_i+\frac{\rho}{2}U_i^2\quad i=2,\ldots,N_j,\label{eq:standard_junction_model}
\end{equation}
Note that $i=1$ is taken as the supplier branch, i.e. the segment that provides most of the flow to the junction.
The second model is denoted \textit{dissipative junction model} (D model), and consists of the junction model proposed in \cite{mynard_2015} which introduces
pressure losses at junctions as follows
\begin{equation}
P_1+\frac{\rho}{2}U_1^2=P_i +\frac{\rho}{2}U_i^2+p_\text{loss}\quad i=2,\ldots,N_j,\label{eq:dissipative_junction_model}
\end{equation}
where 
\begin{equation}
p^\text{loss}_i = K_i \left( \frac{1}{2} \rho u^2_1\right) ,
\end{equation}
with $u_1$ the velocity in the supplier branch. Moreover, the loss coefficient $K_i$ is defined as
\begin{equation}
K_i = \left(2 C_i + \frac{u^2_i}{u^2_1} - 1 \right) \frac{u^2_i}{u^2_1} ,
\end{equation}
with
\begin{equation}
C_i=1-\frac{1}{\lambda_i\psi_i}\cos{\big[\tfrac{3}{4}(\pi-\varphi_i)\big]},
\end{equation}
being $\lambda_i=\frac{Q_i}{Q_1}$, $\psi_i=\frac{A_1}{A_i}$ and $\varphi_i=\frac{3}{4}(\pi-\phi_i)$, where $\phi_i=\theta_1-\theta_i$.
Here, $Q_i$, $A_i$ and $\theta_i$ are the flow rate, the lumen area and the angle measured from the supplier branch, for branch $i$, while  $Q_1$, $A_1$ and $\theta_1$ stand for the same quantities in the supplier branch.
This model was developed for 2D bifurcations. Equations \eqref{eq:masscons}-\eqref{eq:standard_junction_model} or equations \eqref{eq:masscons}-\eqref{eq:dissipative_junction_model} are complemented with Riemann invariants for outgoing waves, resulting in a non-linear system of algebraic equations. For further details see \cite{Muller_1D.HighOrderLTS-2016}.

\subsubsection{Stenosis Model}

Stenoses are accounted for through the lumped parameter model proposed in \cite{young_1973}, for which the pressure drop across the constriction takes the form
\begin{equation}
\Delta P = K_v\frac{\mu}{D}U+K_t\frac{\rho}{2}\bigg[\frac{A}{A_s}-1\bigg]^2 U^2 + K_u\rho L_s \frac{dU}{dt},\label{eq:Young_model}
\end{equation}
where $U$ and $A$ ($D$ the diameter) are the velocity and lumen area in the unobstructed part of the vessel, $L_s$ is the stenosis length, $A_s$ is the minimum stenosis area, and $K_v$, $K_t$ and $K_u$ are model parameters characterizing viscous, turbulent and inertial effects, respectively.

\subsubsection{Numerical method}

Equations \eqref{eq:1Dmodel} with appropriate boundary and coupling conditions are numerically solved using the local time stepping high-order finite volume method presented in \cite{Muller_1D.HighOrderLTS-2016}, where full details of the implementation are given.

\subsection{Automatic Stenosis Detection}\label{sec:detection}

Stenoses are detected in a fully automatic fashion to ensure reproducibility.
A modified version of the algorithm proposed in~\cite{Shahzad_StenosisDetectionAutomaticAlgorithm-2013} to detect stenotic regions in centerlines was proposed and implemented.
With this procedure, the values of $A_s$ and $L_s$ in equation \eqref{eq:Young_model} are characterized for each stenotic lesion.
See Supplementary Material for details.

\subsection{Scenarios and Model Parameters}\label{sec:scenarios}

Common parameters for both models are $\rho=1.05 \text{g/cm}^3$ and $\mu=4\text{ cP}$. For the 3D model, the only fixed parameter is $P_{\text{ref}}=10\text{ mmHg}$. For the 1D model, fixed parameters are $P_0=88\text{ mmHg}$, $K_t=1.52$, $K_u=1.0$ and $\varpi=11$. Patient-specific parameters are: $P_{\text{ao}}$, $R_{\text{out}}$, estimated as explained in \cite{bulant_head_2016}, and $K_v$ which depends on the geometry of the lesion \cite{Seeley_StenoModelParameterEstimations-1976}, that is
\begin{equation}
K_v = 32\dfrac{0.83 L_s + 3.28 \sqrt{A_s}}{2\sqrt{A}}\left[ 0.75 \dfrac{A}{A_s} + 0.25\right]^2.\label{eq:YoungModel.Kv}
\end{equation}
The parameters that play a main role in the viscous dissipation are $\varpi$ and $K_v$. Since the focus of this work is given to the pressure drop predictions delivered by the 1D model, then different scenarios are proposed, with baseline parameters $\varpi^o=11$, as suggested in \cite{hunter_numerical_1972} for coronary flow, and $K_v^o$ determined from equation \eqref{eq:YoungModel.Kv}. These scenarios are
\begin{itemize}
\item \textbf{Raw (R) scenario}: $\varpi^o$, without stenosis models
\item \textbf{Practical (P) scenario}: $\varpi^o$ and $K_v^o$
\item \textbf{Intermediate (I) scenario}: $\varpi^o$ and $K_v=f_K K_v^o$, $f_K$ \textit{estimated}
\item \textbf{Best-case (B) scenario}: $\varpi=f_{\varpi}{\varpi}^o$ and $K_v=f_K K_v^o$, $f_K$ and $f_{\varpi}$ both \textit{estimated}
\end{itemize}

Observe that we created the R scenario in which there are no stenoses in the models. That is, models are purely 1D, and there is no need neither to apply the stenosis detection algorithm nor to define parameters $K_t$, $K_u$, $K_v$. In the P scenario the stenosis detection algorithm is applied, and all parameters are defined uniformly for all stenoses in all patients.
In scenarios I and B, we refer to \textit{estimated} parameters, which implies that data from 3D simulations is extracted and some parameters are identified using a Kalman filter-based data 
assimilation approach (see \cite{Caiazzo2017} for details) to deliver the best possible match. Factor (or stenosis factor) $f_K$ is estimated to define the value of $K_v$ for each stenosis in each vascular network such that the pressure drop $\Delta P$ delivered by the stenosis in the 1D model matches the pressure drop obtained in that stenosis from the 3D model.
Factor (or profile factor) $f_{\varpi}$ is estimated to define the value of ${\varpi}$, which defines the velocity profile for each vascular network, such that the pressure at outlet locations in the 1D model matches the pressure obtained in these outlets from the 3D simulation.
For both estimates ($f_K$ and $f_{\varpi}$), the cost functionals used are the time-averaged errors along a single cardiac cycle.

Scenarios I and B are reported because they feature the best achievable results in terms of 1D modeling. In fact, parameters in these cases are stenosis-specific ($K_v$) and patient-specific ($\varpi$) and constructed to match data from 3D simulations.

These four scenarios are combined with the two junctions models: S (standard, see equation \eqref{eq:standard_junction_model}) and D (dissipative, see equation \eqref{eq:dissipative_junction_model}) for a total of eight scenarios, denoted by $\text{Y}_\text{X}$, $\text{Y}\in\{\text{R},\text{P},\text{I},\text{B}\}$ and $\text{X}\in\{\text{S},\text{D}\}$.

\subsection{Data Analysis and Comparisons}

The value of \FFRd{3} and \FFRd{1} predicted by the different scenarios are compared at four relevant locations in each network. These locations correspond to major vessels (left anterior descending, circumflex, ramus intermedius and right coronary arteries) at locations $\ell_{4\text{P}}=\{\frac{\ell}{4},\frac{\ell}{2},\frac{3\ell}{4},\ell\}$, where $\ell$ is the total length of the vessel. Also, the location of invasive \FFR measurement is denoted by $\ell_{\text{\FFR}}$.

Comparisons are reported using mean and standard deviation of the values and discrepancies, Bland-Altman analysis ($m_\text{BA}\pm\text{SD}_\text{BA}$). Pearson's correlation coefficient $r$, as well as coefficients of the linear approximation $\text{\FFRd{1}}=a \text{\FFRd{3}} + b$ are also reported. A cut-off value of $\text{\FFRd{3}}=0.8$ is used to identify functional stenoses, and taking \FFRd{3} as ground truth, the prevalence (Prev), and classification indexes such as the area under the receiver operating characteristic curve (AUC), accuracy (Acc), sensitivity (Sen), specificity (Spe), positive predictive value (PPV) and negative predictive value (NPV) are calculated for each one of the different 1D scenarios considered in this work.

\section{Results}\label{sec:Results}

Table~\ref{tb:stenoses_stats} presents the statistical comparison between the stenotic pressure drop in the 3D model and the stenotic pressure drops as predicted by all the 1D scenarios described in Section~\ref{sec:scenarios}.
The statistics of the stenosis morphology defined by the automatic algorithm explained in Section~\ref{sec:detection} is also reported, where $\frac{A_s}{A}$ is the severity and $L_s$ the stenosis length, see equations \eqref{eq:Young_model} and \eqref{eq:YoungModel.Kv}. Also, the Reynolds number computed in the 3D model is given as the average between the inlet and outlet Reynolds numbers. In the cases of scenarios $\text{I}$ and $\text{B}$, the statistics of the value of the stenosis factor $f_K$ estimated using the Kalman filter is given. Finally, in the $\text{B}$ scenario, the statistics of the profile factor $f_{\varpi}$ estimated using the Kalman filter is reported.

Table~\ref{tb:FFR_stats} reports the statistical analysis of the results at $\ell_{4\text{P}}$ and $\ell_{\text{\FFR}}$ locations, summarizing the performance of the eight scenarios involving 1D models with respect to the ground truth prediction of \FFRd{3}.

Figure~\ref{fig:FFR_corr_BA} displays the scatter plot and the Bland-Altman plots for the comparison between \FFRd{3}, the gold standard, and \FFRd{1} as given by the eight different scenarios.

\section*{Discussion}\label{sec:Discussion}

\begin{figure}[ht]
\begin{subfigure}[ ]{.49\textwidth}
	\centering 
\includegraphics[scale=0.32]{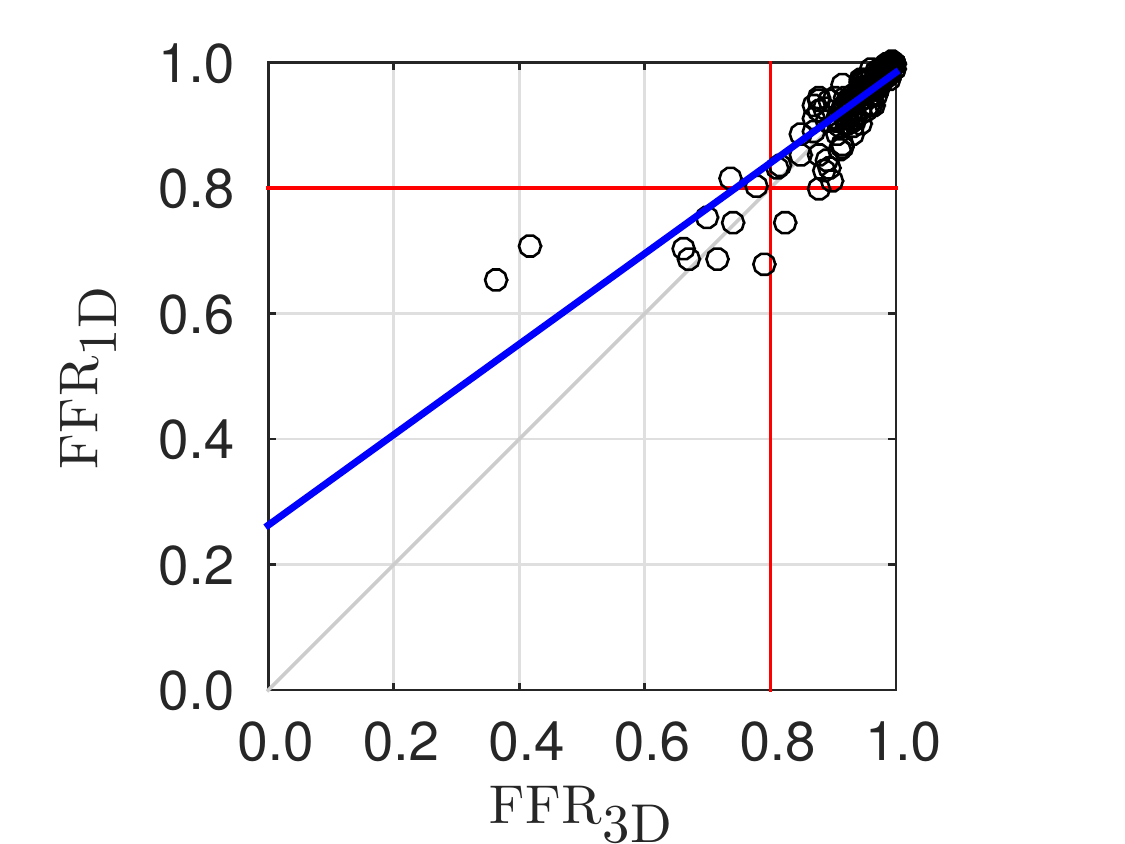}\includegraphics[scale=0.32]{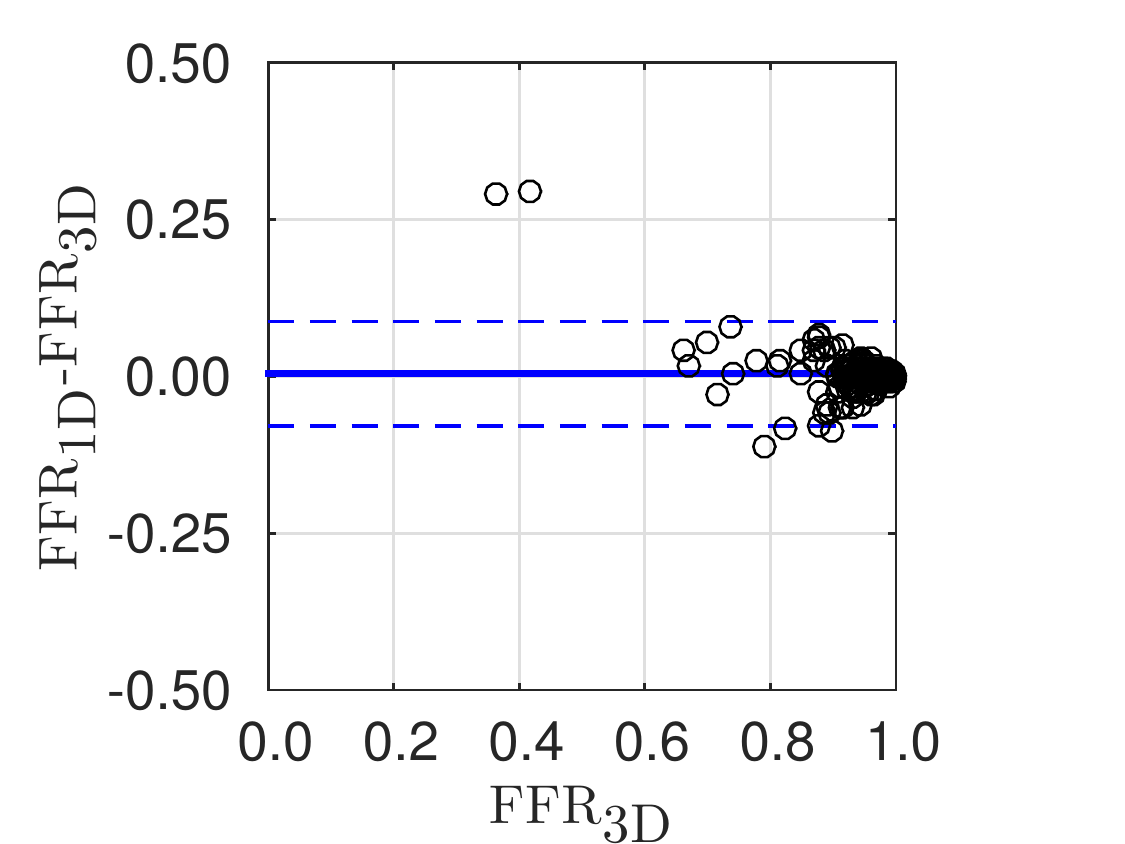}
 	\caption{Scenario $\text{R}_\text{D}$}
\end{subfigure}
\begin{subfigure}[ ]{.49\textwidth}
	\centering
	\includegraphics[scale=0.32]{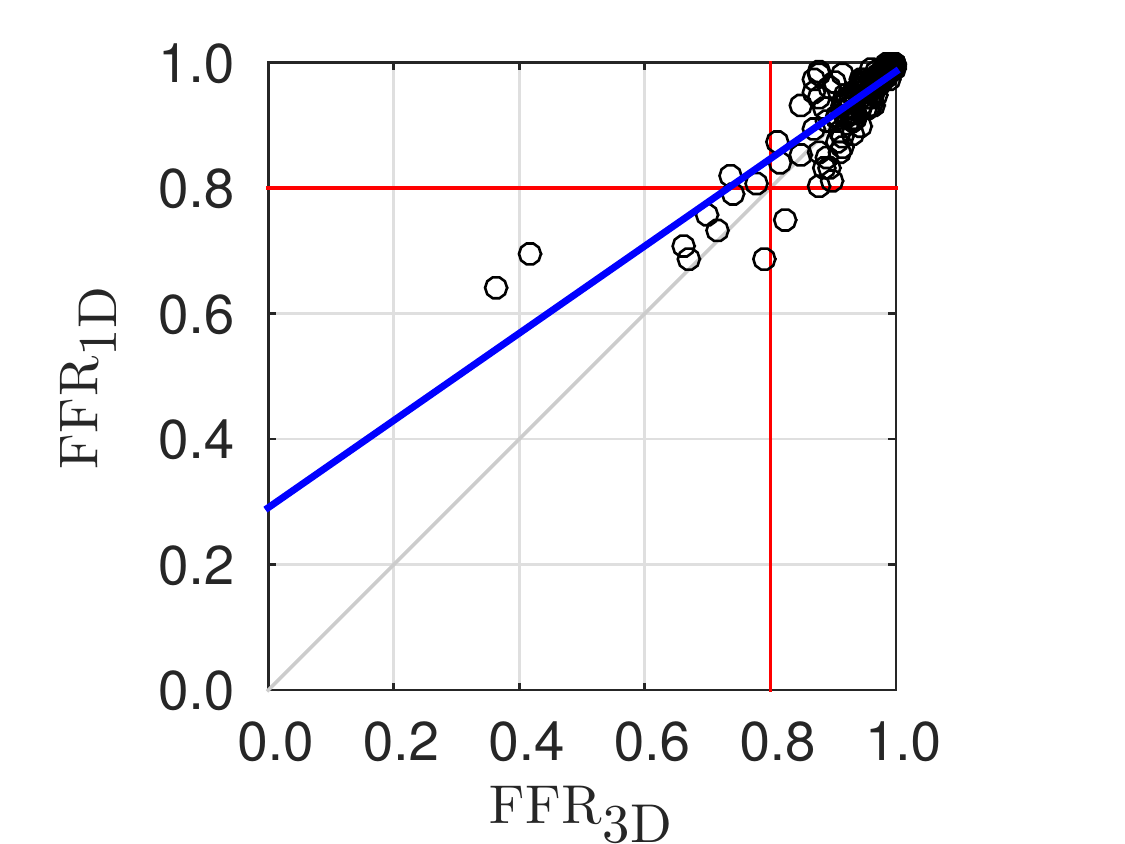}\includegraphics[scale=0.32]{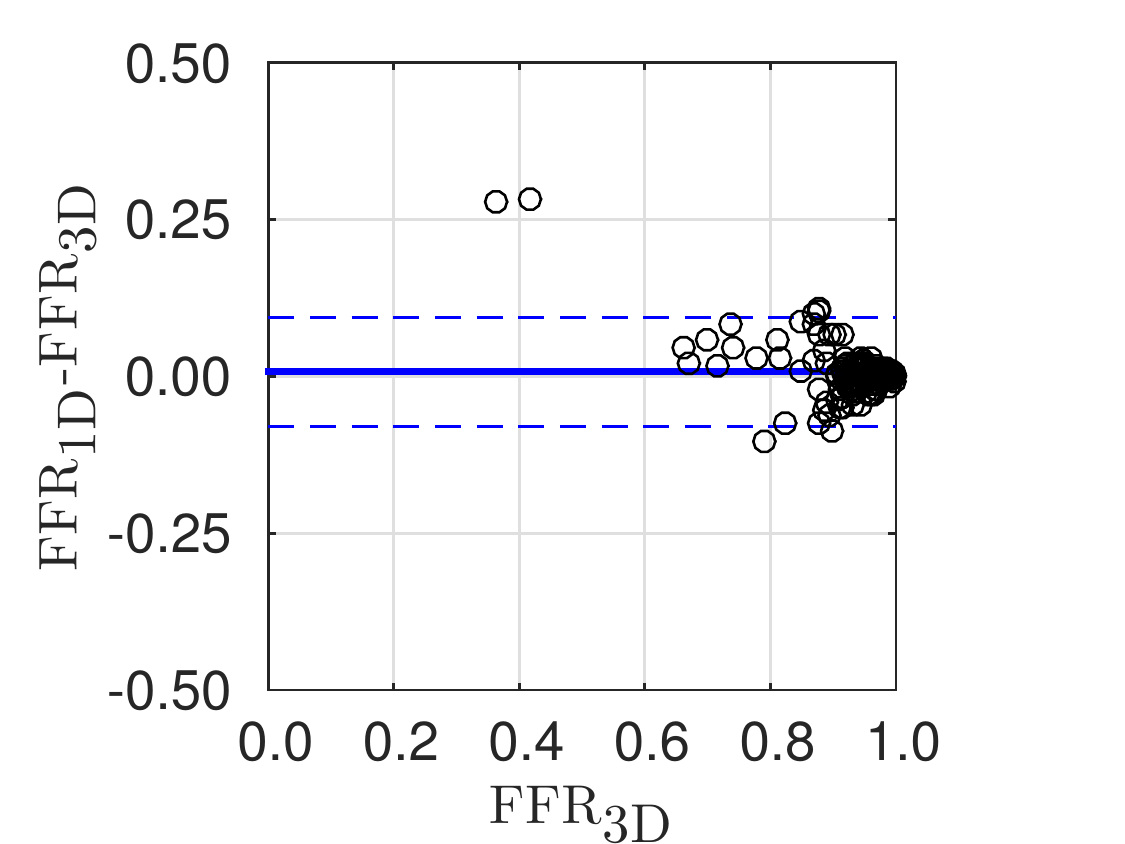}
 	\caption{Scenario $\text{R}_\text{S}$}
\end{subfigure}
\begin{subfigure}[ ]{.49\textwidth}
	\centering
	\includegraphics[scale=0.32]{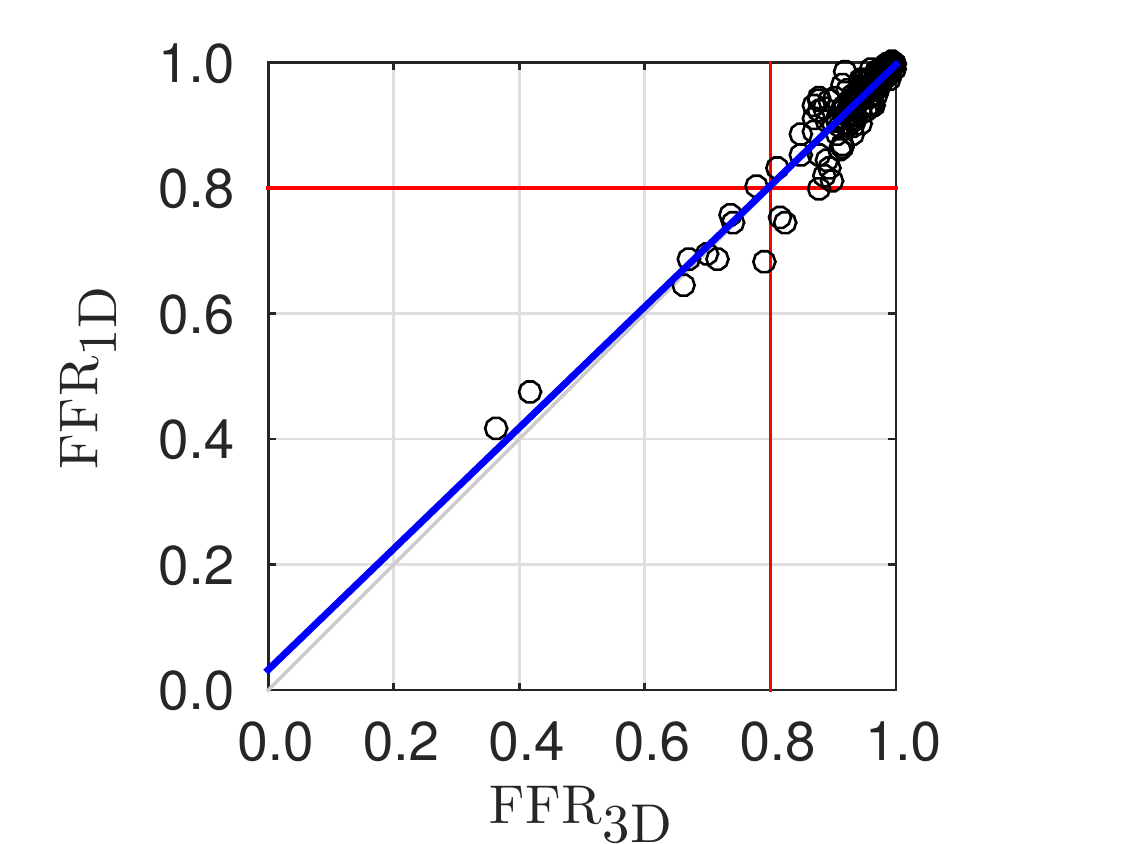}\includegraphics[scale=0.32]{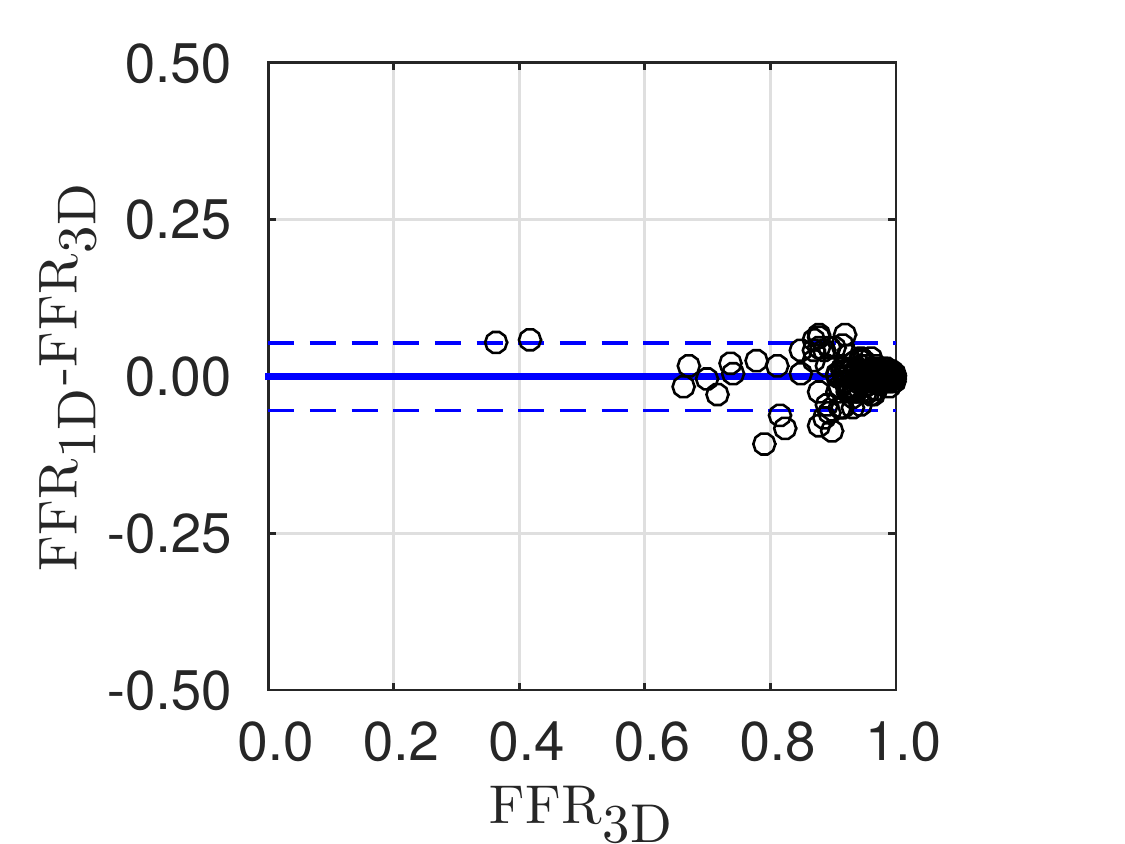}
 	\caption{Scenario $\text{P}_\text{D}$}
\end{subfigure}
\begin{subfigure}[ ]{.49\textwidth}
	\centering
	\includegraphics[scale=0.32]{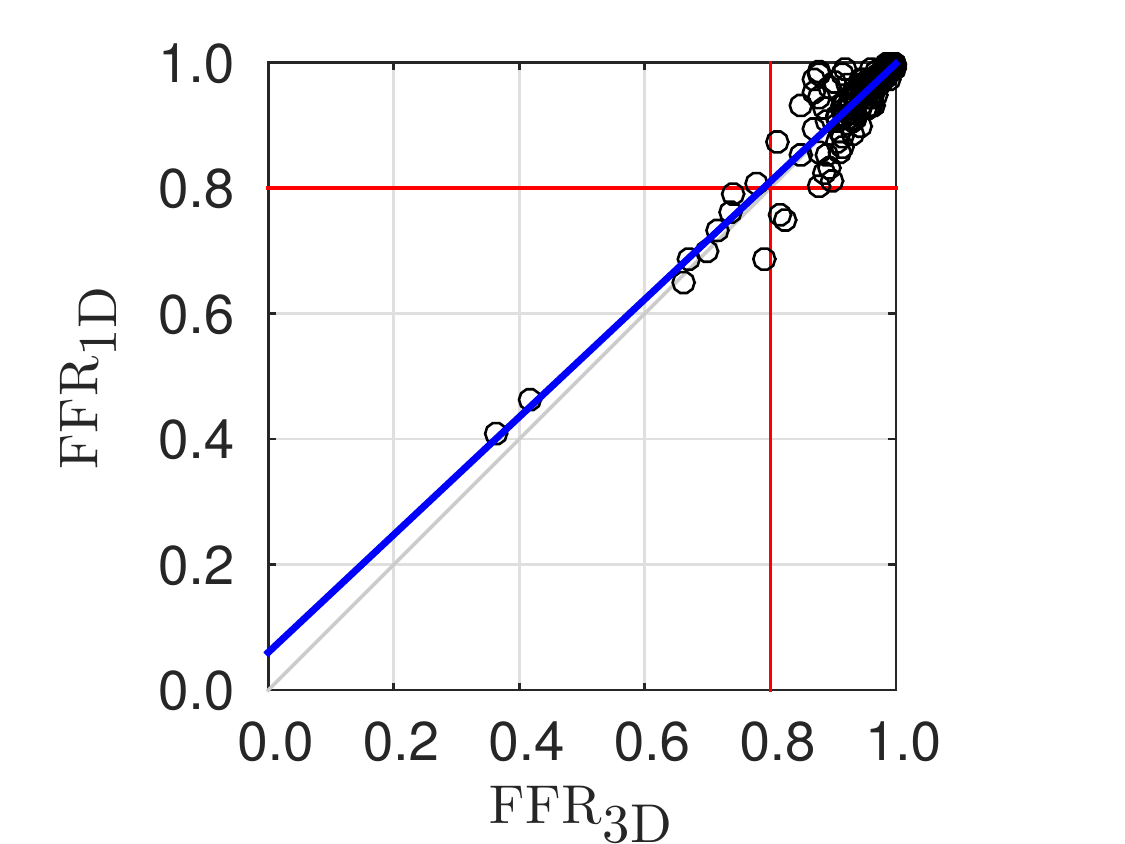}\includegraphics[scale=0.32]{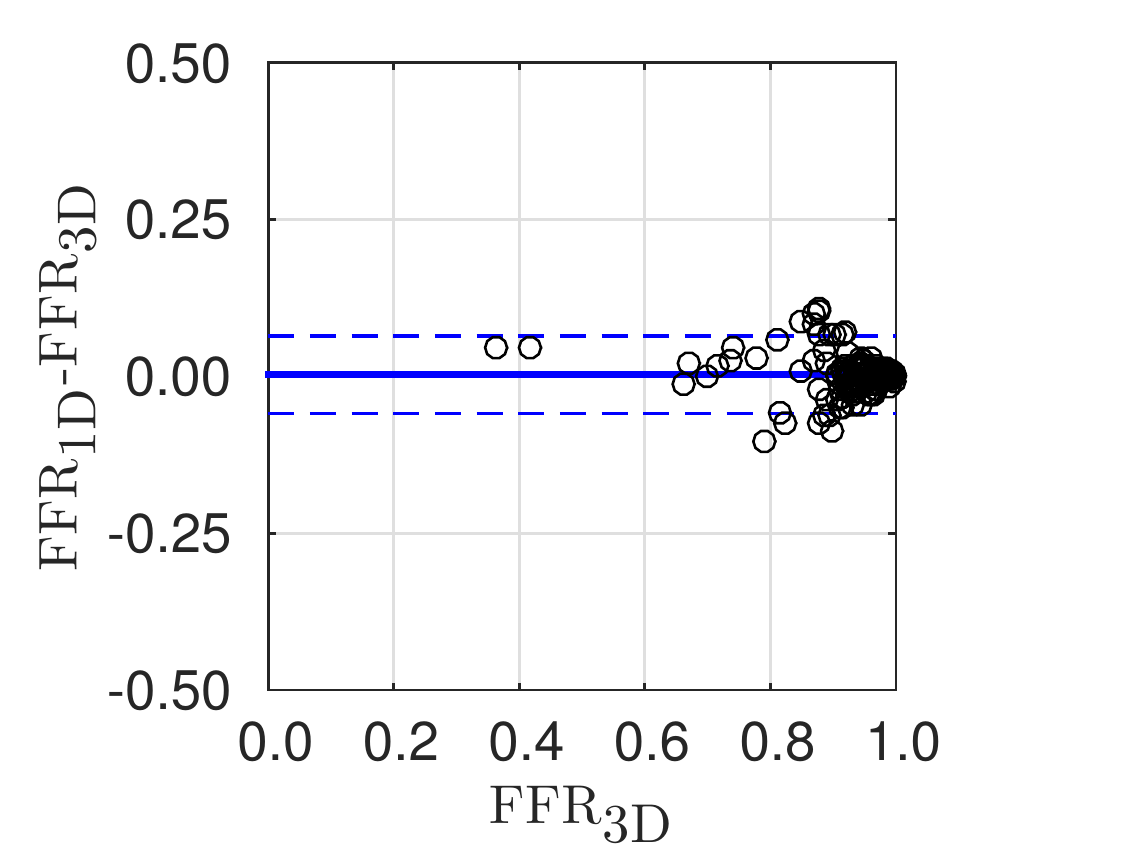}
 	\caption{Scenario $\text{P}_\text{S}$}
\end{subfigure}
\begin{subfigure}[ ]{.49\textwidth}
	\centering
	\includegraphics[scale=0.32]{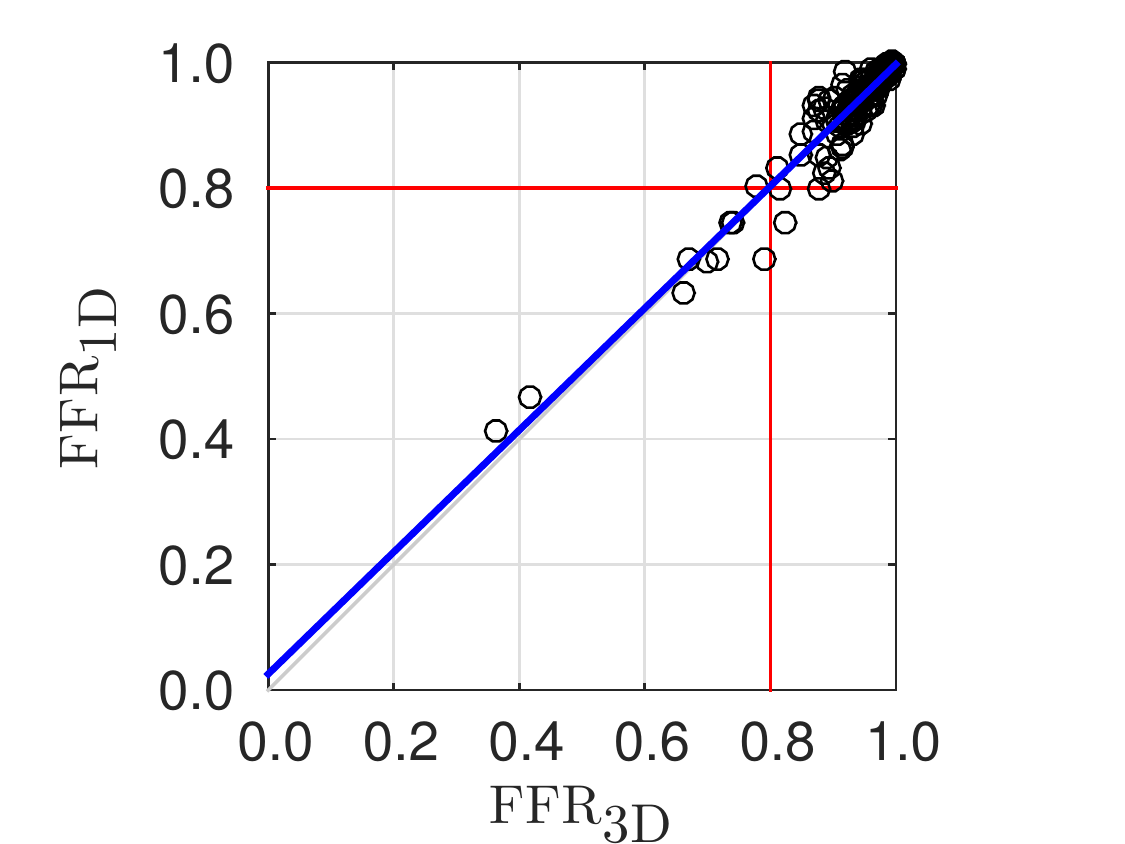}\includegraphics[scale=0.32]{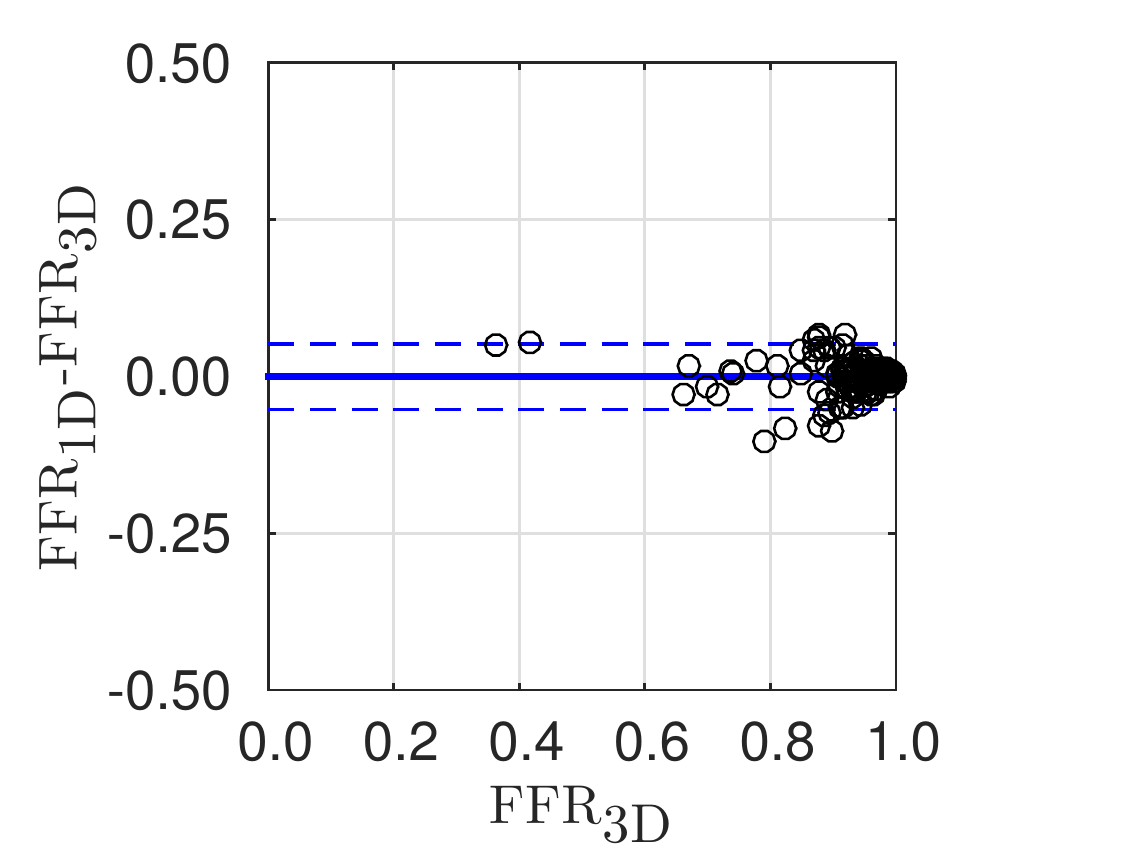}
 	\caption{Scenario $\text{I}_\text{D}$}
\end{subfigure}
\begin{subfigure}[ ]{.49\textwidth}
	\centering
	\includegraphics[scale=0.32]{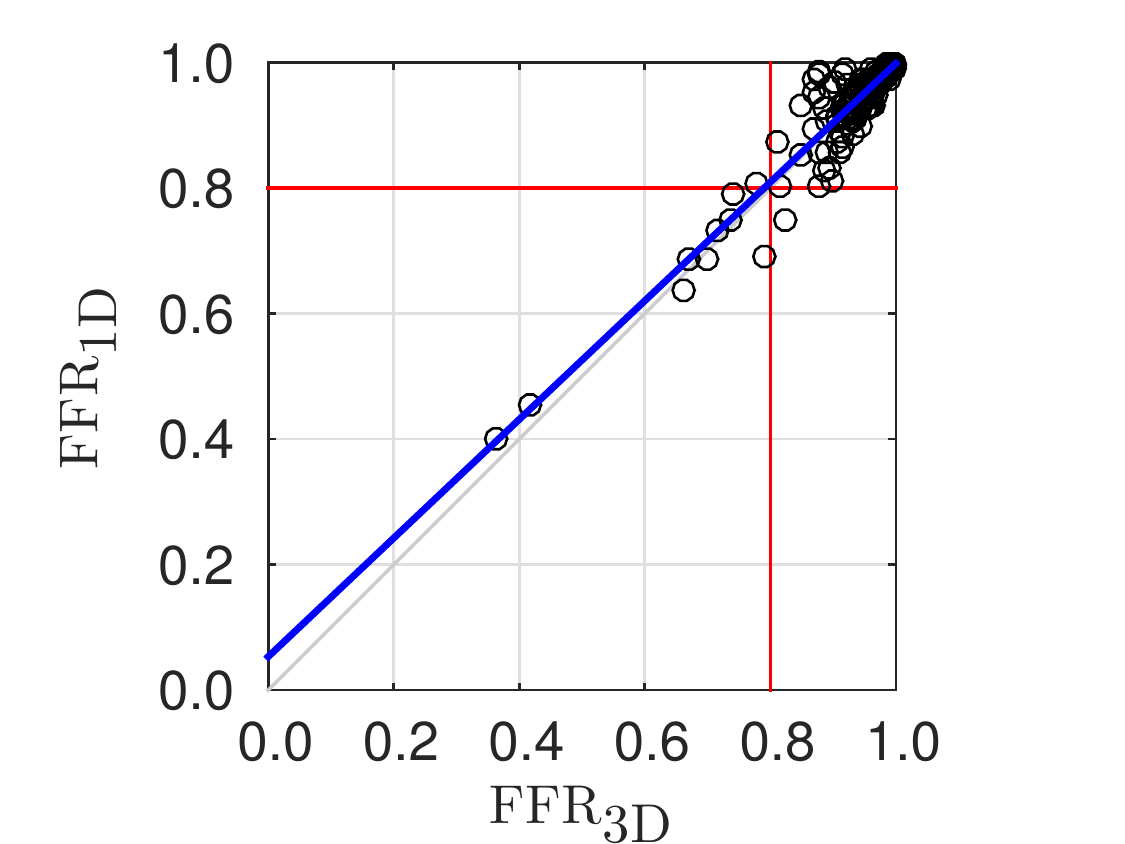}\includegraphics[scale=0.32]{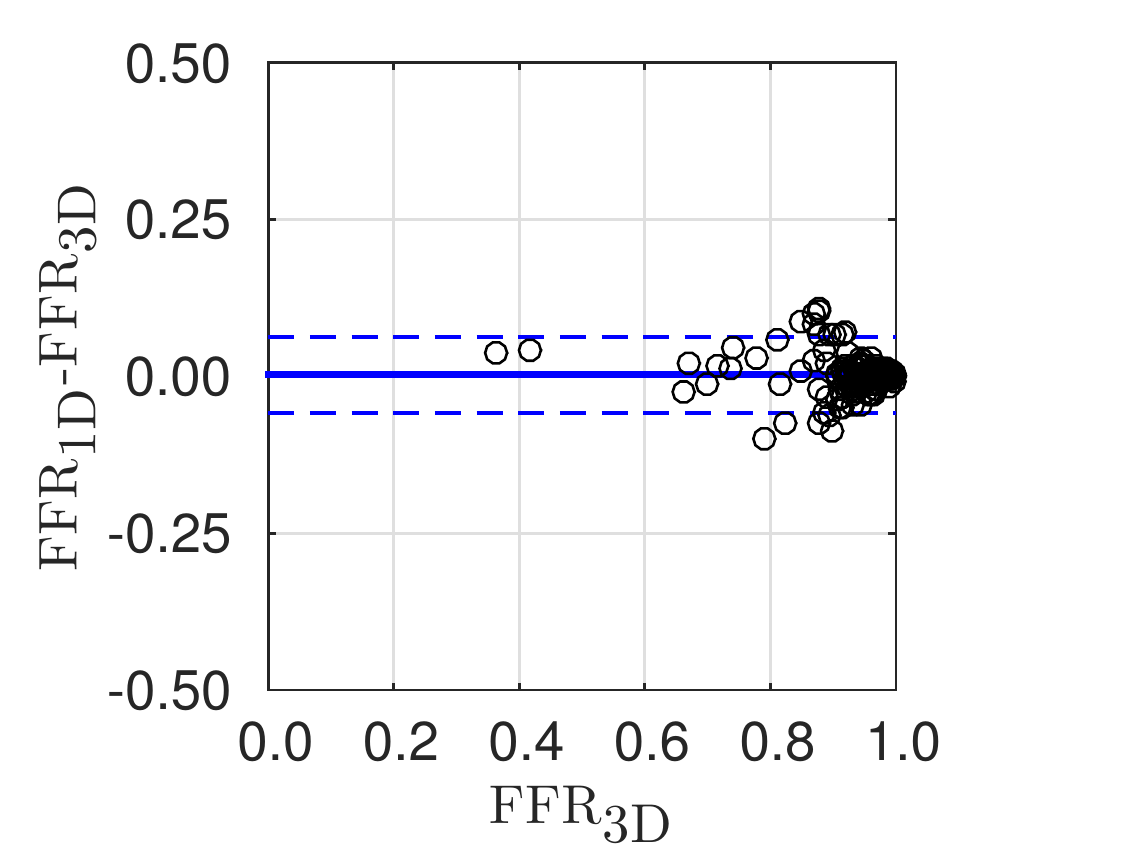}
 	\caption{Scenario $\text{I}_\text{S}$}
\end{subfigure}
\begin{subfigure}[ ]{.49\textwidth}
	\centering
	\includegraphics[scale=0.32]{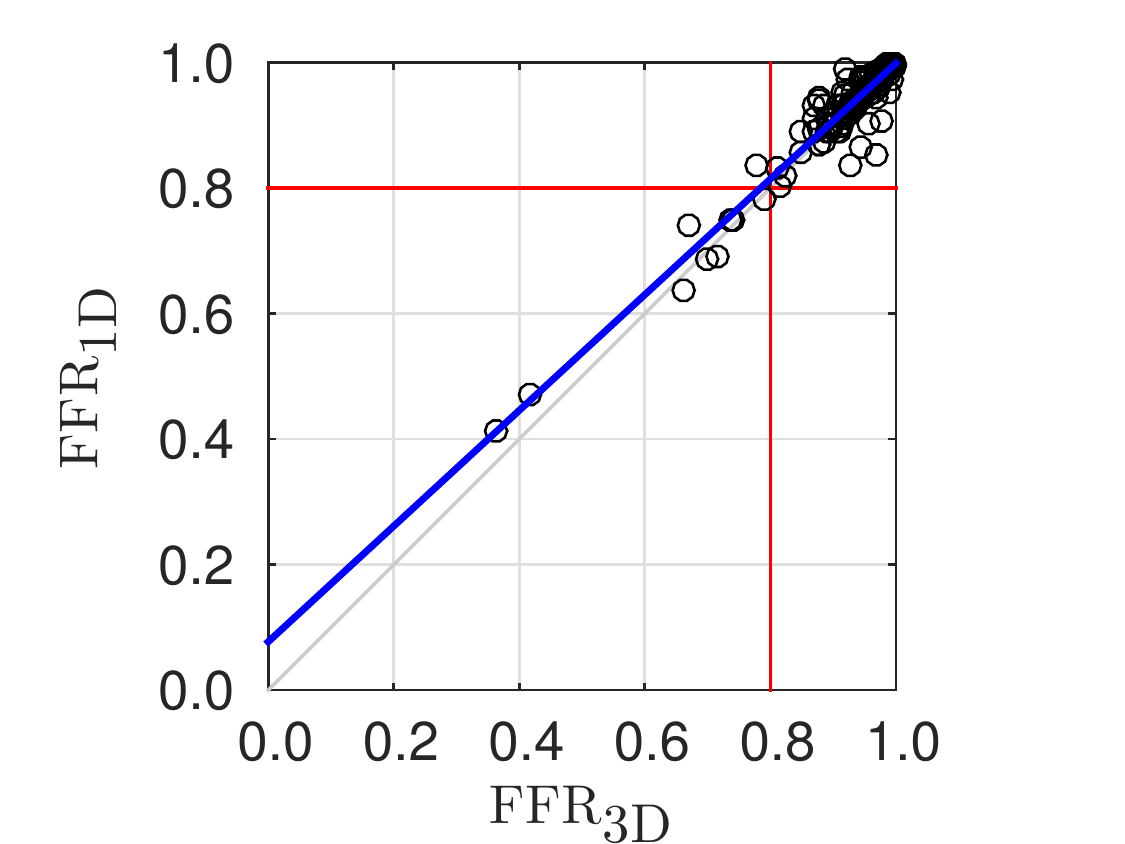}\includegraphics[scale=0.32]{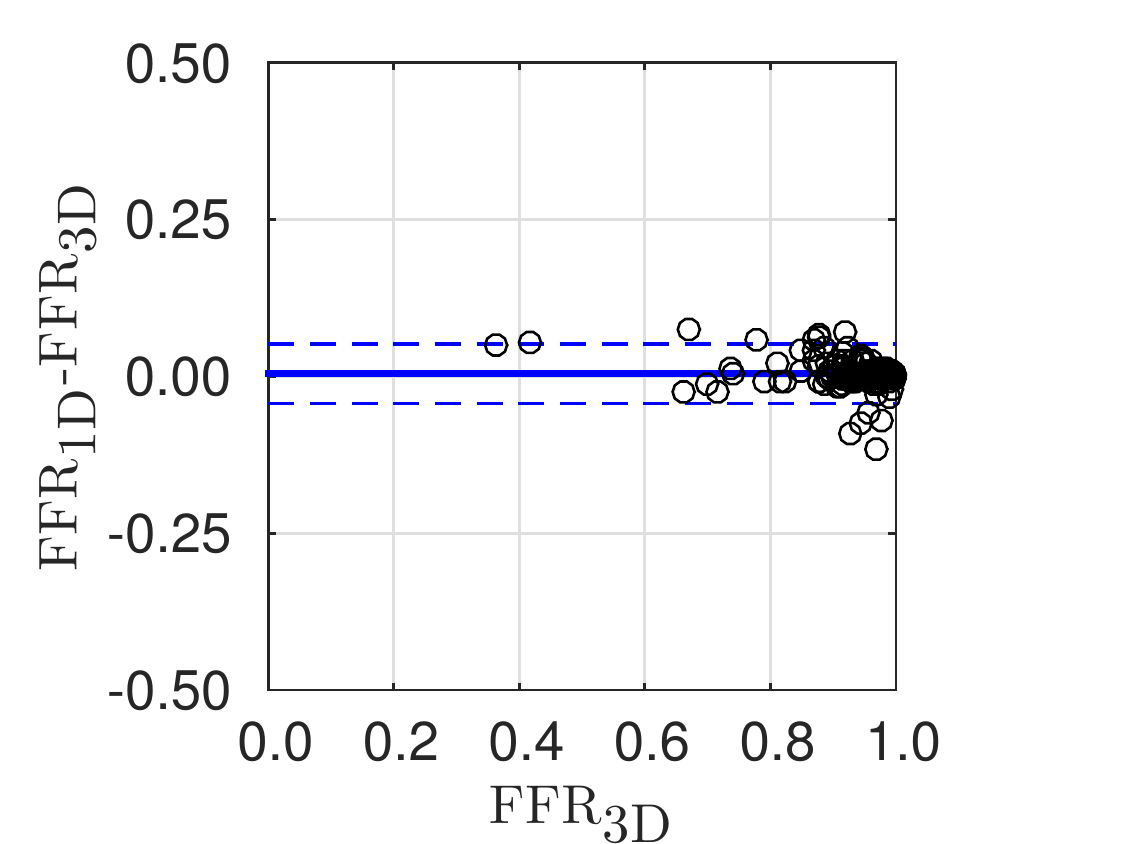}
 	\caption{Scenario $\text{B}_\text{D}$}
\end{subfigure}
\begin{subfigure}[ ]{.49\textwidth}
	\centering
	\includegraphics[scale=0.32]{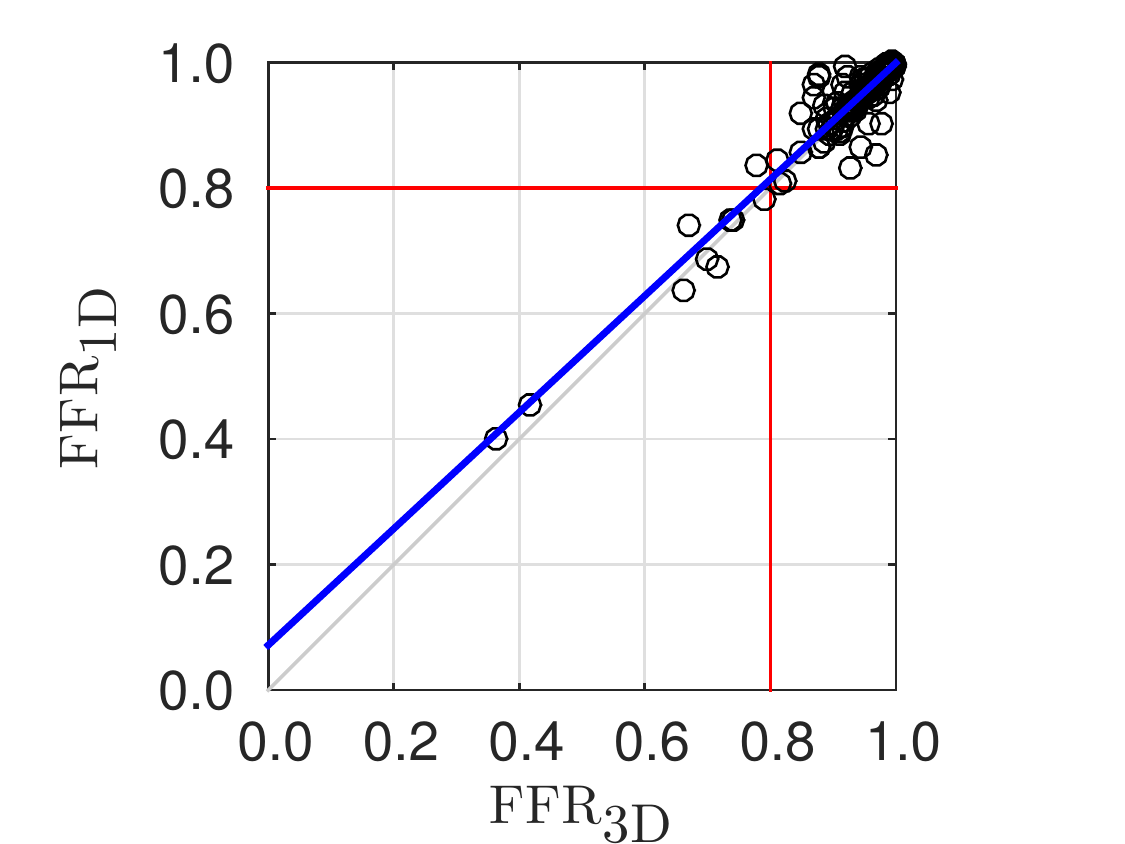}\includegraphics[scale=0.32]{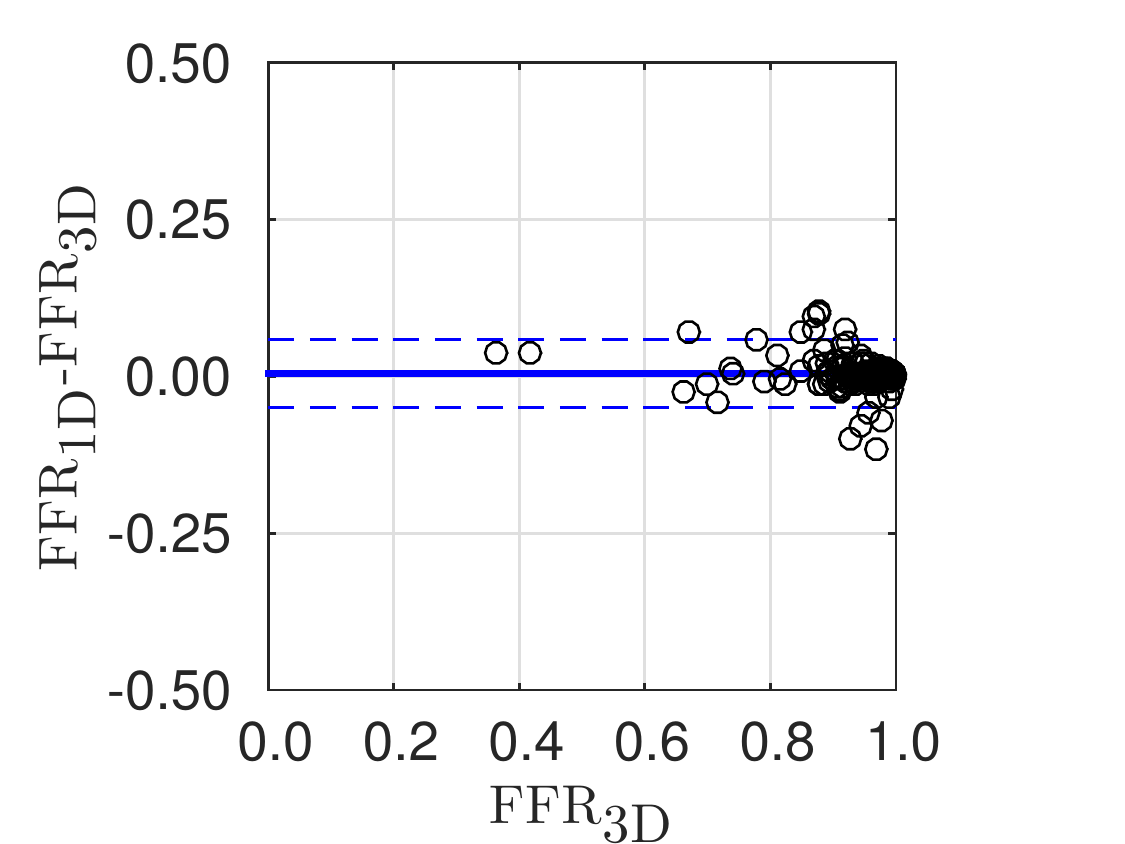}
 	\caption{Scenario $\text{B}_\text{S}$}
\end{subfigure}

\caption{Scatter and Bland-Altman plots featuring comparison between the gold standard \FFRd{3} and \FFRd{1} for different scenarios, 
$\text{Y}_\text{X}$, $\text{Y}\in\{\text{R},\text{P},\text{I},\text{B}\}$ and $\text{X}\in\{\text{S},\text{D}\}$, with
$\text{R}$: raw, $\text{P}$: practical, $\text{I}$: intermediate, $\text{B}$: best, $\text{S}$: standard junction and $\text{D}$: dissipative junction.
Results correspond to four locations, $\ell_{4\text{P}}$, in the interrogated vessels.}
\label{fig:FFR_corr_BA}
\end{figure}

Inspecting the results reported in Table~\ref{tb:FFR_stats} at the four locations $\ell_{4\text{P}}$, we observe that the all 1D model scenarios provide excellent classification capabilities when compared to the 3D model. Moreover, almost no bias and a small standard deviation are obtained for the difference $\text{\FFRd{1}}-\text{\FFRd{3}}$. Overall, considering the junctions as dissipative (model D), instead of standard (model S), does not bring substantial improvements neither to the correlation coefficients ($a,b,r,m_{\text{BA}},\text{SD}_{\text{BA}}$) nor to the classification indexes (AUC, Acc, Sen, Spe, PPV, NPV). More in detail, we note that the plain 1D model (scenarios $\text{R}_{\text{D}}$ or $\text{R}_{\text{S}}$) provides the poorest correlation with the 3D model (coefficients of linear regression $a=0.72$, $b=0.26$,  $r=0.88$ for $\text{R}_\text{D}$). By adding the stenosis model with a one-fits-all strategy for the parameter calibration (scenarios $\text{P}_{\text{D}}$ or $\text{P}_{\text{S}}$), the correlation coefficients significantly improve ($a=0.96$, $b=0.03$,  $r=0.95$ for $\text{P}_\text{D}$). This can also be appreciated in Figure~\ref{fig:FFR_corr_BA}, where the alignment of the point cloud around the $45^\circ$ line is clear. In such plot, the correction of some outliers is also noticeable. As a consequence, the use of stenosis models is mandatory to have the best the 1D modeling realm.

Moreover, notwithstanding the setting of stenosis-specific parameters tuned with 3D data brings some improvements, there is no considerable gain in the correlation coefficients ($a=0.97$, $b=0.03$,  $r=0.96$ for $\text{I}_\text{D}$), while classification indexes remain invariant. Even in the best case scenario in which we further estimate the velocity profile such that 1D terminal pressures match those of 3D models, there is no relevant improvement in the model capabilities. The stenosis parameter estimated from 3D data resulted in average very close to the unit value (see Table~\ref{tb:stenoses_stats}) which was taken in the one-fits-all approach, and as suggested in the original contribution \cite{young_1973}.

It is important to stress that the comparisons presented here exclusively focuse on the ability of 1D models to predict pressure drops $\Delta P$ in stenotic lesions. This has been achieved by setting the same flow rate boundary conditions extracted from the 3D models to the 1D counterparts, guaranteeing the consistency in the flow regime in among the models. The discrepancy in the stenotic $\Delta P$ estimation is also observed in Table~\ref{tb:stenoses_stats}. While the plain 1D model (scenarios $\text{R}_{\text{D}}$ or $\text{R}_{\text{S}}$) underestimates the $\Delta P$, the inclusion of stenosis models yields larger $\Delta P$, rendering the alignment in the correlation line seen in Figure~\ref{fig:FFR_corr_BA}. On the one hand, regarding the lesion-specific parameter $K_v$ estimated from 3D data (scenarios I and B), it is remarkable that, in average, it results very close to the unit value (i.e. $f_K= 0.97\pm0.51$), close to the value chosen for the one-fits-all approach in scenario P (theoretically $f_K=1$). On the other hand, regarding the characterization of the velocity profile given by network-specific parameter $\varpi$, the estimation using 3D data indicates that there is room for improvement (i.e. $f_\varpi=0.74\pm0.16$, model D) in order to improve the selection criterion for such parameter.

Analyzing the prediction of \FFR in clinically relevant locations $\ell_{\text{FFR}}$ (see Table~\ref{tb:FFR_stats}), we observe that both the correlation coefficients and the classification indexes continue to be excellent, and the AUC index grows almost to a perfect unitary value. Particularly, the only low value is obtained for the PPV, which is in part a consequence of the low prevalence sample. However, in the best case scenarios (scenarios $\text{B}_{\text{D}}$ and $\text{B}_{\text{S}}$) demonstrate that, when properly tuned, 1D models provide an exact match with 3D models in term of diagnostic capabilities. Furthermore, at these clinically relevant locations, since in average they are more distally placed, cumulative effect of upstream junctions causes the dissipative junction (scenarios D) to outperform the standard junction (scenarios S). Therefore, our recommendation is to employ this junction model.

The relatively low prevalence of positive \FFRd{3} values in the $\ell_{4\text{P}}$ population, equal to 0.06, might raise concern as whether this fact is being favorable to validating our working hypothesis on the ability of 1D simulations to match 3D results. In order to rule out such concern we considered an alternative population $\hat{\ell}_{4\text{P}}$ in which we only included the points of $\ell_{4\text{P}}$ for vessels in which at least one of its points had a \FFRd{3}$<0.85$. In that case, the number of sampled points is 36, with a prevalence of 0.28. Statistics for all scenarios are consisted between populations $\ell_{4\text{P}}$ and $\hat{\ell}_{4\text{P}}$. Particularly, results for scenario $\text{P}_\text{D}$ are:  mean difference of $0.01\pm0.04$, significant correlation of 0.96, slope of 0.97 and intercept of 0.03 for the linear regression and overall AUC, accuracy, sensitivity, specificity, positive predictive value, and negative predictive value of 0.98, 0.92, 0.90, 0.92, 0.82, 0.96. Comparison of these results with the ones reported in Table \ref{tb:FFR_stats} for the same scenario show that considerations about the accuracy of 1D simulations for FFR prediction remain valid.

In a nutshell, practical scenarios making use of one-fits-all stenosis parameters are sufficient for an excellent estimation of the pressure drop occurring as predicted by 3D simulations. The use of dissipative junctions is not mandatory, although it slightly improves the capabilities of \FFRd{1}, with more gains at clinically relevant locations.

Moreover, 1D models only require the use of workstations which are easily available in the medical facilities, while 3D models necessarily make use of clusters of high performance computers which poses an unrealistic scenario in clinical routine.

\section*{Final Remarks}\label{sec:FinalRemarks}

The results reported in this work indicate that \FFRd{1} simulations can be reliable surrogates of \FFRd{3} models to assess functional significance of coronary stenoses. Even if 1D models must be endowed with stenoses models to effectively predict pressure drops in lesions, a one-fits-all strategy to set up stenoses parameters rendered excellent predictive capabilities in terms of classification indexes when regarding the \FFRd{3} as the ground truth. Adopting such practical strategy, and adding dissipative junctions, when compared to \FFRd{3}, \FFRd{1} renders a mean difference of $0.00\pm0.03$ ($-0.02\pm0.05$), and overall accuracy, sensitivity, specificity, positive predictive value, and negative predictive value of 0.98, 0.90, 0.99, 0.82, and 0.99 (0.97, 1.00, 0.97, 0.75, and 1.00), respectively, to detect significant stenoses at several locations in the coronary network (at clinically relevant locations).

Remarkably, it has been possible to show that, in the best case scenario conceived by setting properly calibrated stenoses parameters, the \FFRd{1} perfectly matched the classification given by \FFRd{3}.

Even if the present results have been obtained for a exploratory sample of patients, the results constitute a first validation survey to test the hypothesis that inexpensive \FFRd{1} can be reliably used instead of \FFRd{3} simulations. As a matter of fact, \FFRd{1} has the benefit of requiring substantial less time and computational resources than \FFRd{3}, which makes this approach affordable in clinical routine. Having demonstrated that the \FFRd{1} concept is feasible, the next step consists in translating this \FFRd{1} tool into the clinic by performing clinical studies.




\section*{Acknowledgements}

Acknowledgements should be brief, and should not include thanks to anonymous referees and editors, or effusive comments. Grant or contribution numbers may be acknowledged.

\section*{Author contributions statement}

Must include all authors, identified by initials, for example:
P.J.B., C.A.B., L.O.M. and R.A.F. conceived the experiment(s),  C.G.B. and P.A.L. collected the data, C.A.B. and G.D.M.T. processed the data, C.A.B. and L.O.M. conducted the experiment(s), P.J.B., C.A.B., L.O.M. and R.A.F. analysed the results. All authors reviewed the manuscript.

\section*{Additional information}

\noindent\textbf{Data availability statement:} The data that support the findings of this study are available on request from the corresponding author [PJB]. The data are not publicly available due to privacy restrictions of research participants.

\noindent\textbf{Competing financial interests:} The authors do not have competing financial interests with regards to the study and results presented in this work.

\FPeval{\size}{2.85}
\FPeval{\sizesm}{1.8}
\FPeval{\xjump}{2}
\FPeval{\yjump}{1.8}

\setlength{\tabcolsep}{1.mm}

\begin{table*}[ht]
	\centering
	{\scriptsize
		\begin{tabular}{cccccc}
			\toprule
			Junction	& $\Delta P$ $[\text{mmHg}]$ & \multicolumn{4}{c}{1D scenarios $\Delta P$ $[\text{mmHg}]$}   \\
			Model & 3D & R$^\dag$: raw & P$^\dag$:practical & I$^\dag$: intermediate & B: best \\
			\midrule
			& 	&	&	&	& \\
			D & \multirow{2}{*}{6.75$\pm$9.08}	&	5.35$\pm$4.29	&	5.52$\pm$6.58	&	5.25$\pm$6.54 & 5.16$\pm$6.56 \\
			S & &	5.34$\pm$4.28	&	5.52$\pm$6.58	&	5.25$\pm$6.54 & 5.18$\pm$6.55 \\
			\bottomrule
		\end{tabular}\vspace{2mm}
		\begin{tabular}{cccccc}
			\toprule
			Junction	& \multirow{2}{*}{$\frac{A_s}{A}$} & \multirow{2}{*}{$L_s$ $[\text{cm}]$} & \multirow{2}{*}{$\Reynolds$} & $f_K$ & $f_{\varpi}$  \\
			Model &  & & & scen. $\text{I},\text{B}$& scen. $\text{B}$\\
			\midrule
			& 	&	& &	& \\
			D & \multirow{2}{*}{0.50$\pm$0.13}	& \multirow{2}{*}{0.55$\pm$0.37}	&	\multirow{2}{*}{185$\pm$94}	&	0.97$\pm$0.51	& 0.74$\pm$0.16\\
			S & &	&	&	0.97$\pm$0.51	&	0.76$\pm$0.22\\
			\bottomrule
		\end{tabular}
	}
	\caption{Comparison of stenotic pressure drop ($\Delta P$) between the 3D model and all 1D scenarios (for both junction models,  D: dissipative and S: standard), in $[\text{mmHg}]$, for all stenoses.
		Marker $^\dag$ indicates scenarios with $p>0.05$ in the paired U-Test, meaning that no significant differences between $\Delta P$ of 1D and 3D predictions was found.
		$\frac{A_s}{A}$: stenosis degree, $L_s$: stenosis length; $\text{Re}$: Reynolds number; $f_K$: stenosis factor estimated by the Kalman filter; $f_{\varpi}$: velocity factor estimated by the Kalman filter.
		Note that $f_\varpi$ statistics are computed over $n=6$ computational models for which at least one stenoses was detected. The remaining statistics were computed using $n=15$ stenosis elements.
	}
	\label{tb:stenoses_stats}
\end{table*}

\setlength{\tabcolsep}{1.5mm}

\begin{table*}[ht]
	\centering
	{\scriptsize
		\begin{tabular}{llccccccccccc}
			\toprule
			& \multicolumn{2}{c}{\multirow{2}{*}{Scenario}}	& \multicolumn{2}{c}{Linear approx.} & Corr. & $\text{\FFRd{1}}-\text{\FFRd{3}}$ & \multicolumn{6}{c}{Prediction value of \FFRd{1} vs. \FFRd{3}}\\
			& 	&  & $a$	& $b$ 	& $r$ & $m_{\text{BA}} \pm \text{SD}_{\text{BA}}$ & AUC & Acc& Sen & Spe & PPV & NPV   \\
			\midrule

\parbox[t]{2mm}{\multirow{16}{*}{\rotatebox[origin=c]{90}{Location of comparison}}}	&
\parbox[t]{2mm}{\multirow{8}{*}{\rotatebox[origin=c]{90}{$\ell_{4\text{P}}$ ($n=156$)}}}	&
			$\text{R}_\text{D}$		&	0.72	&	0.26	&	0.88	&	0.00$\pm$0.04$^\dag$	&	0.97	&	0.98	&	0.80	&	0.99	&	0.89	&	0.99\\
&		&	$\text{P}_\text{D}$		&	0.96	&	0.03	&	0.95	&	0.00$\pm$0.03$^\dag$	&	0.97	&	0.98	&	0.90	&	0.99	&	0.82	&	0.99\\
&		&	$\text{I}_\text{D}$		&	0.97	&	0.03	&	0.96	&	0.00$\pm$0.03$^\dag$	&	0.97	&	0.98	&	0.90	&	0.99	&	0.82	&	0.99\\
&		&	$\text{B}_\text{D}$		&	0.92	&	0.08	&	0.96	&	0.00$\pm$0.02$\,\,\,$	&	0.96	&	0.99	&	0.90	&	1.00	&	1.00	&	0.99\\
&		&	$\text{R}_\text{S}$		&	0.69	&	0.29	&	0.87	&	0.01$\pm$0.04$^\dag$	&	0.97	&	0.98	&	0.80	&	0.99	&	0.89	&	0.99\\
&		&	$\text{P}_\text{S}$		&	0.94	&	0.06	&	0.94	&	0.00$\pm$0.03$^\dag$	&	0.97	&	0.98	&	0.90	&	0.99	&	0.82	&	0.99\\
&		&	$\text{I}_\text{S}$		&	0.95	&	0.05	&	0.94	&	0.00$\pm$0.03$^\dag$	&	0.98	&	0.99	&	0.90	&	0.99	&	0.90	&	0.99\\
&		&	$\text{B}_\text{S}$		&	0.93	&	0.07	&	0.95	&	0.00$\pm$0.03$\,\,\,$	&	0.95	&	0.99	&	0.90	&	1.00	&	1.00	&	0.99\\
\\ \cline{2-13} \\
&	\parbox[t]{2mm}{\multirow{8}{*}{\rotatebox[origin=c]{90}{$\ell_{\text{FFR}}$ ($n=32$)}}}	&
			$\text{R}_\text{D}$		&	0.70	&	0.26	&	0.80	&	-0.01$\pm$0.07$\,\,\,$	&	0.99	&	0.97	&	1.00	&	0.97	&	0.75	&	1.00\\
&		&	$\text{P}_\text{D}$		&	1.01	&	-0.03	&	0.92	&	-0.02$\pm$0.05$\,\,\,$	&	0.99	&	0.97	&	1.00	&	0.97	&	0.75	&	1.00\\
&		&	$\text{I}_\text{D}$		&	1.02	&	-0.04	&	0.92	&	-0.02$\pm$0.05$\,\,\,$	&	0.99	&	0.97	&	1.00	&	0.97	&	0.75	&	1.00\\
&		&	$\text{B}_\text{D}$		&	1.02	&	-0.03	&	0.92	&	-0.01$\pm$0.05$^\dag$	&	1.00	&	1.00	&	1.00	&	1.00	&	1.00	&	1.00\\
&		&	$\text{R}_\text{S}$		&	0.69	&	0.28	&	0.79	&	-0.01$\pm$0.07$^\dag$	&	0.99	&	0.94	&	0.67	&	0.97	&	0.67	&	0.97\\
&		&	$\text{P}_\text{S}$		&	1.00	&	-0.01	&	0.91	&	-0.02$\pm$0.05$^\dag$	&	0.99	&	0.94	&	0.67	&	0.97	&	0.67	&	0.97\\
&		&	$\text{I}_\text{S}$		&	1.01	&	-0.03	&	0.91	&	-0.02$\pm$0.06$^\dag$	&	0.99	&	0.94	&	0.67	&	0.97	&	0.67	&	0.97\\
&		&	$\text{B}_\text{S}$		&	1.04	&	-0.04	&	0.93	&	-0.01$\pm$0.05$^\dag$	&	1.00	&	0.97	&	0.67	&	1.00	&	1.00	&	0.97\\
			\bottomrule
		\end{tabular}
	}
	\caption{
	Statistical results of predictive capabilities of \FFRd{1} when compared with \FFRd{3} for the different scenarios
	$\text{Y}_\text{X}$, $\text{Y}\in\{\text{R},\text{P},\text{I},\text{B}\}$ and $\text{X}\in\{\text{S},\text{D}\}$, with
	$\text{R}$: raw, $\text{P}$: practical, $\text{I}$: intermediate, $\text{B}$: best, $\text{S}$: standard junction and $\text{D}$: dissipative junction.
	Values of \FFR compared at four locations, $\ell_{4\text{P}}$, in the interrogated vessel as well as at the clinically relevant location for diagnosis, $\ell_{\text{\FFR}}$.
	Sample sizes are obtained from the 29 computational models.
	Prevalence of functional stenoses according to \FFRd{3}: $0.06$ for $\ell_{4\text{P}}$ and $0.09$ for $\ell_{\text{FFR}}$.
	Linear approximation coefficients defined by $a$ and $b$.
	$r$: Pearson's correlation coefficient ($p<0.05$ for all models).
	$m_{\text{BA}}\pm\text{SD}_{\text{BA}}$: mean and standard deviation of Bland-Altman analysis for the difference $\text{\FFRd{1}}-\text{\FFRd{3}}$.
	Marker $^\dag$ indicates correlation ($p\geq0.05$) between 1D and 3D models.
	Predicted values (AUC, Acc, Sen, Spe, PPV, NPV) computed using \FFRd{3} as gold standard and a cut-off value of $\text{FFR}\geq0.8$.}
	\label{tb:FFR_stats}
\end{table*}

\end{document}